\documentclass[aps,prb,twocolumn,showpacs]{revtex4}

\usepackage{graphicx}

\begin{document}

\title{\bf Effects of Resonant Cavity on Macroscopic Quantum Tunneling\\
of Fluxon in Long Josephson Junctions}
\author{Ju H. Kim and Ramesh P. Dhungana$^*$}
\affiliation{
Department of Physics and Astrophysics, University of North Dakota,
Grand Forks, ND 58202-7129}

\begin{abstract}
We investigate the effects of high-$Q_c$ resonant cavity on macroscopic quantum tunneling
(MQT) of fluxon both from a metastable state to continuum and from one degenerate
ground-state of a double-well potential to the other.  By using a set of two coupled
perturbed sine-Gordon equations, we describe the tunneling processes in linear long
Josephson junctions (LJJs) and find that MQT in the resonant cavity increases due to
potential renomalization, induced by the interaction between the fluxon and cavity.
 Enhancement of the MQT rate in the weak-coupling regime is estimated by using the experimantally accessible range of the model parameters.  The tunneling rate from the
metastable state is found to increase weakly with increasing junction-cavity interaction
strength.  However, the energy splitting between the two degenerate ground-states of the
double-well potential increases significantly with increasing both the interaction
strength and frequency of the resonant cavity mode.  Finally, we discuss how the resonant
cavity may be used to tune the property of Josephson vortex quantum bits.
\end{abstract}


\maketitle

\section{introduction}

Experimentally observed\cite{Wa} quantum behavior of Josephson vortices (i.e., fluxons) at
ultra-low temperatures has opened up a possibility of realizing quantum computers based on
long Josephson junctions (LJJs).   This observation led to much interest on Josephson
vortex quantum bit\cite{Cl,KDP,oJVQ} (qubit) as an alternative to the previously proposed
superconducting qubits.  Similar to other approaches based on Josephson junctions such as
charge,\cite{CQ} phase,\cite{PQ} and flux\cite{FQ} qubits, Josephson vortex qubit (JVQ) is
also a promising candidate for quantum computation application.  Due to its weak
interaction with decoherence sources in the environment at low temperatures, the JVQ may
have significant advantages over the other superconducting qubits.  For instance, a
significantly longer decoherence time was suggested as one such advantage.\cite{KDP}

The JVQ takes advantage of the coherent superposition of two spatially separated states
arising from the low temperture property of a trapped fluxon in a double-well potential.
This property includes (i) energy quantization and (ii) macroscopic quantum
tunneling\cite{Wa}(MQT).  We note that, for linear LJJs, the fluxon potential for either
metastable state or JVQ may be obtained\cite{Fab} by using Nb-AlO$_x$-Nb junctions and by
implanting either one or two microresistors in the insulator layer, respectively.  For
application of JVQs, tuning both the decoherence time and the level of entaglement by
controlling the qubit property is essential.  However, due to its weak interaction with
external perturbations, an effective tuning mechanism for JVQ is less clear.  Recent
studies\cite{mc1,mc2} on using microwave cavity for both tuning a single phase qubit and
inducing interaction between either two charge or two phase qubits suggest that resonant
cavity may be used for JVQ to serve the same purpose.

Earlier studies on the effects of resonant cavity indicate\cite{ES,TS} that both electric
and magnetic fields of the cavity couple to the Josephson junction since the cavity
electromagnetic (EM) mode behaves similar to a phonon mode\cite{Pho} which interacts with
the fluxon.  The effects of resonant cavity on the fluxon dynamics in LJJ
stacks\cite{Sakai,KP,Kl} have been studied both experimentally\cite{Exp1,Exp2} and
theoretically.\cite{Sim1,Sim2,SSN}  These studies show that when the coupling between LJJ
and resonant cavity is spatially uniform, no force is exerted on the fluxon by the cavity,
but its dynamics may become modified.  These studies suggest that the interaction between
LJJ and a resonant EM wave mode of the cavity promotes\cite{Pete} collective dynamics of
fluxons.  The in-phase locking mode of the fluxon dynamics is shown to be
enhanced\cite{Pete} by the cavity EM mode.

These studies also suggest that the junction-cavity interaction may be used to change the
qubit property.   The property of JVQ depends on MQT between two spatially separated
states of the fluxon.  We note that MQT represents quantum particle-like collective
exciations.\cite{KI,Sh}  As semi-classical theories indicate that the MQT rate\cite{KM}
depends on the potential barrier height, the JVQ can be tuned by adjusting the
potential-well for the fluxon.  This adjustment can be achieved by potential
renormalization induced by the junction-cavity interaction since this interaction can
strongly affect the fluxon tunneling processes, similar to phonon assisted tunneling in
Josephson junctions.\cite{Mak}  We note that a two-level atom interacting with a quantized
radiation field, described by the Jaynes-Cummings model,\cite{JC} is also similar to the
JVQ-cavity system that we consider in the present work.  The potential renormalization for
fluxon suggests that the resonant cavity may be used as a tool for controlling the JVQ
property.  As the fluxon tunneling processes may be controlled externally by tuning either
the junction-cavity coupling strength or the resonant frequency, the effects of the
resonant cavity depend on the nature of the interaction.  However, the influence of
junction-cavity interaction on the MQT rate has not been understood clearly.

In this paper, we investigate the effects of the junction-cavity both on MQT from
metastable state and on the ground-state energy splitting in a double-well potential.  We
note that, to focus on the interaction between LJJ and a single resonant cavity mode, we
consider only a high-$Q_c$ cavity.  First, we estimate the MQT rate for the fluxon in a
single LJJ and for the phase-locked fluxons in a coupled LJJ stack by computing the local
and non-local contributions.  Then, we estimate the effects of resonant cavity on the JVQ
property by computing the ground-state energy splitting.   Before proceeding further, we
outline the main result.  (i) {\it The potential barrier for a fluxon in the metastable
state is not affected by increasing neither the junction-cavity interaction nor the
resonant frequency of the cavity EM mode}.  (ii) {\it The non-local contribution to the
tunneling rate due to the junction-cavity interaction is negligible in the weak-coupling
regime.} (iii) {\it Due to potential renormalization induced by the junction-cavity
interaction, the potenital barrier height for the fluxon trapped in a double-well
potential is reduced.  This reduction leads to increase in the ground-state energy
splitting for the JVQ with increasing junction-cavity coupling and resonant frequency}.

The outline of the remainder of the paper is as follows. In Sec. II, we describe the
LJJ-cavity system by using a set of two perturbed sine-Gordon equations.  In Sec. III, the effects of resonant cavity on the fluxon tunneling rate from the metastable state in a LJJ
are discussed.  In Sec. IV, we discuss MQT of phase-locked fluxons from the
metastable state in a vertical stack of two coupled LJJs.  In Sec. V, the effects of interaction between LJJ and a single mode in high-$Q_c$ cavity on JVQ are estimated by computing the ground-state energy splitting.  Finally, we summarize the result
and conclude in Sec. VI.

\section{Coupled long Josephson junctions in resonant cavity}

To examine i) one-fluxon tunneling in a single LJJ, ii) phase-locked two-fluxon tunneling
in a stack of two coupled LJJs, and iii) the ground-state energy splitting in JVQ, we
start with coupled perturbed sine-Gordon equations\cite{Sakai} for describing two LJJs
which interact with resonant cavity\cite{TS}
\begin{eqnarray}
{\partial^2 \over \partial x^2}
\bigg ( \varphi_{1} - {\cal S} \varphi_{2} \bigg ) -
{\partial^2 \varphi_{1} \over \partial t^2} -
\sin \varphi_{1} = {\cal F}_1
\label{sGlay1}  \\
{\partial^2 \over \partial x^2}
\bigg ( \varphi_{2} -{\cal S} \varphi_{1} \bigg ) -
{\partial^2 \varphi_{2} \over \partial t^2} -
\sin \varphi_{2} = {\cal F}_2
\label{sGlay2}
\end{eqnarray}
where $x$ and $t$ are the dimensionless coordinates in units of $\lambda_J
\gamma^{-1}({\cal S})$ and $\omega_p^{-1}$, respectively.  Here $\gamma^{-1}
({\cal S})= \sqrt{1-{\cal S}^2}$ and $\omega_p$ denotes the plasma frequency.  The dynamic
variable $\varphi_{i}$ represents the difference between the phase $\phi$ of the
superconductor order parameter for the two superconductor (S) layers $i$ and $i-1$ (i.e.,
$\varphi_i = \phi_i - \phi_{i-1}$).  The strength of magnetic induction coupling between
two LJJs is denoted by $\cal S$.  Here we set $\hbar =k_B=c=1$ for convenience.  The
perturbation term $\cal F$ of for each LJJ which is given by
\begin{equation}
{\cal F}_i =\beta {\partial \varphi_{i} \over \partial t} +
f_i -g_E {d^2q_r \over dt^2}- \epsilon_i \delta(x-x_i^o)
\sin\varphi_{i}
\label{perturb}
\end{equation}
accounts for the contribution from dissipation ($\beta$), bias current ($f=J^B/J_c$),
resonant cavity ($g_E$), and microresistors ($\epsilon= (J_c-J'_c)l_b/J_c\lambda_J$).  Here
$x_i^o$, $J^B$, $J_c$, $J'_c$, $l_b$ ($\ll \lambda_J$) and $\lambda_J$ denote the position
of microresistors in the insulator layer of the $i$-th junction, the bias current density,
the critical current density, the modified current density, the length of the LJJ in which
$J_c$ is modified, and the Josephson length, respectively.  We note that dissipation, bias
currents, resonant cavity and microresistors on the phase dynamics lead to different effects.

We account for the perturbation contribution due to resonant cavity by following Tornes
and Stroud\cite{TS} and by assuming that the cavity supports a single harmonic oscillator
mode which may be represented by the {\it displacement} variable $q_r$ as
\begin{equation}
{d^2 q_r \over dt^2} + {\omega_r \over Q_c}{dq_r \over dt} +
\omega_r^2 q_r= {g_E \gamma({\cal S})\over M_{osc}} \int dx
{\partial^2 \over \partial t^2}
\bigg ( \varphi_{1} + \varphi_{2} \bigg ) .
\label{sGlay3}
\end{equation}
Here $\omega_r$, $Q_c$, and $M_{osc}$ are the dimensionless oscillator frequency in units
of $\omega_p$, the cavity quality factor, and the "mass" of the oscillator mode,
respectively.  For simplicity, {\it we neglect the second term on the left hand side of
Eq. (\ref{sGlay3}) by assuming that the cavity is non-dissipative (i.e., high-$Q_c$
cavity).}  Also, we assume that the cavity electric field $\bf E$ is uniform within the
junction by considering the spatially uniform junction-cavity coupling $g_E$ of
\begin{equation}
g_E = -{\epsilon_d \over 2e} \sqrt{M_{osc} \over
4\pi}~{\bf{E} \cdot \hat{\bf z}} ,
\end{equation}
where $\epsilon_d$ is the dielectric constant.  As we will discuss below, the position
independent coupling $g_E$ does not change the fluxon motion directly but yields
potential renormalization when a microresistor is present.

To estimate the effects of interaction between LJJ and resonant cavity analytically, we
consider the weak perturbation $\cal F$ limit.  As each perturbation term in Eq.
(\ref{perturb}) is small and does not change the form of the kink solution within the
lowest order approximation,\cite{MS} we describe the fluxon motion in terms of the center
coordinate $q(t)$.   In the absence of both the perturbation terms (${\cal F}=0$) and the
magnetic induction effect (${\cal S}=0$), the fluxon solution to Eq. (\ref{sGlay1}) is
given by
\begin{equation}
\varphi_{i}(x,t) \approx 4 \tan^{-1}
\left[ e^{\gamma (v_i ) [x-q_i(t)]}
\right]~,
\label{soliton}
\end{equation}
in the non-relativistic limit (i.e., $v \ll 1$).  Here $q_i(t)=v_i t$ denotes the center
coordinate for the fluxon, and $v$ is the fluxon speed in units of Swihart velocity.
Equation (\ref{soliton}) represents propagation of nonlinear wave as a ballistic particle.
The perturbation contributions of $\cal F$ only affect the dynamics of fluxon expressed in
the $q$ coordinate.

We now describe the fluxon phase dynamics  in the coupled LJJ using the center coordinate
$q_i$ representation.  The energy of the fluxon may be seen easily from the Euclidean
Lagrangian (i.e., $\tau=it$),
\begin{equation}
{\cal L}= {\cal L}_o + {\cal L}_{mag} + {\cal L}_{pert} +
{\cal L}_{osc} + {\cal L}_{coup}~.
\label{Lagran}
\end{equation}
The first three terms for $\cal L$ of Eq. (\ref{Lagran}) describe the LJJ contributions,
while the remaining two terms arise from the resonant cavity.  First, we discuss the LJJ
contributions to Lagrangian ${\cal L}$.  The unperturbed part of LJJ is described by the
Lagrangian ${\cal L}_o$ given by
\begin{equation}
{\cal L}_o =\sum_i \int {dx \over 2} \bigg[ \bigg(
{\partial \varphi_i \over \partial \tau} \bigg)^2 +
\bigg({\partial\varphi_i \over \partial x} \bigg)^2 +
2(1-\cos\varphi_i)\bigg]~.
\end{equation}
The Lagrangian contribution from the magnetic induction effect, ${\cal L}_{mag}$, is
given by
\begin{equation}
{\cal L}_{mag} = {\cal S} \int dx \left(
{\partial \varphi_1 \over \partial x} \right) \left(
{\partial \varphi_2 \over \partial x} \right)~.
\end{equation}
We note that ${\cal L}_{mag}$ accounts for the interaction energy $E_{int}$ between two
LJJs due to the magnetic induction effect.  The perturbation contribution to the
Lagrangian, ${\cal L}_{pert}={\cal L}_{nd}+{\cal L}_{d}$, is expressed as the sum of two
terms: i) the non-dissipative (${\cal L}_{nd}$) and ii) dissipative (${\cal L}_{d}$) part.
The non-dissipative contribution comes from the bias currents and microresistors.  The
non-dissipative Lagrangian ${\cal L}_{nd}$ is expressed as the sum of the contributions
from the bias current (${\cal L}_{bias}$) and microresistors (${\cal L}_{pin}$) (i.e.,
${\cal L}_{nd} = {\cal L}_{bias} + {\cal L}_{pin}$).  The bias current contribution
${\cal L}_{bias}$ is given by
\begin{equation}
{\cal L}_{bias}= \sum_i \int dx f_i \varphi_i~,
 \end{equation}
and the inhomogeneity contribution due to microresistors ${\cal L}_{pin}$ is given by
\begin{equation}
{\cal L}_{pin}=\sum_i\int dx~ \epsilon_i \delta(x-x_i^o)
(1-\cos\varphi_i)~.
\end{equation}
We note that ${\cal L}_{pin}$ accounts for the fluxon pinning energy $E_{pin}$.  These
non-dissipative contributions provide the bare fluxon potential $V(q)$.  On the other
hand, the dissipative Lagrangian ${\cal L}_d$ accounts for the interaction between the
fluxon and environment.  The effects of this contribution may be described\cite{CL} by
following Caldeira and Leggett and by representing the environment as a heat bath.  The
heat bath is represented as harmonic oscillators with generalized momenta $P_i$ and
coordinates $Q_i$.  The dissipation Lagrangian ${\cal L}_d$ which accounts for the
coupling between the phase ($\varphi$) and oscillator ($Q_i$) variables is given by
\begin{equation}
{\cal L}_{d}=\int dx \sum_{i} \left[ {P_i^2 \over 2m_i} +
{m_i\omega_i^2 \over 2} \left( Q_i - {c_i \varphi \over m_i
\omega_i^2} \right)^2 \right]~.
\end{equation}
Here, the spectral function $J_\beta(\omega)$,
\begin{equation}
J_\beta(\omega) = {\pi \over 2} \sum_i
{c_i^2 \over m_i \omega_i^2} ~
\delta(\omega - \omega_i) = \beta\omega ,
\label{Spec}
\end{equation}
is used to reproduce the dissipation effects ($\beta$) in Eq. (\ref{perturb}).  The effects
of dissipation on a two-state system has been studied extensively by using the spin-boson
model.\cite{LCDFGZ}  In the adiabatic approximation, the energy splitting for the two-state
system is known to be reduced\cite{LCDFGZ} in the dissipative environment.  However, this
result does not\cite{CL} imply that the effects of the interaction between the two-state
system and a single oscillator, which represents either a phonon or quantized radiation
field, on the energy spliting is similar.  In our discussion below, we neglect the
dissipation effects by setting $\beta = 0$ since these effects are small at low
temperatures, and we focus on the effects due to a resonant cavity.

We now discuss the high-$Q_c$ resonant cavity contribution to the Lagrangian ${\cal L}$ of
Eq. (\ref{Lagran}).  The resonant cavity is modeled by using Lagrangian for a single
harmonic oscillator which represents a single EM-mode supported by the cavity.  The
Lagrangian for this single mode oscillator ${\cal L}_{osc}$ is written as
\begin{equation}
{\cal L}_{osc}={M_{osc} \over 2} \bigg( {dq_r \over
d\tau} \bigg)^2 + {K \over 2}~q_r^2 ,
\end{equation}
where $K$ is the "spring constant" and $q_r$ denotes the oscillator coordinate.  We note
that the oscillator frequency $\omega_r$ in Eq. (\ref{sGlay3}) is given by
$\omega_r=(K/M_{osc})^{1/2}$.  The capacitive coupling between LJJ and resonant cavity is
described by the Lagrangian ${\cal L}_{coup}$ as
\begin{equation}
{\cal L}_{coup}= - g_E \bigg( {dq_r \over d\tau} \bigg)~\int dx~
\sum_i \bigg( {\partial \varphi_i \over \partial \tau} \bigg)~.
\label{coup}
\end{equation}
Here we assume that the coordinate $q_r$ is spatially homogeneous and focus on the effects
of the uniform $\bf E$-field in the cavity.  We note that the interaction between LJJ and
resonant cavity yields the non-local effects, similar to those from the dissipation term
(i.e., $\beta \neq 0$).

We estimate MQT of fluxon by using the usual semiclassical approach\cite{GJS} of starting
with the partition function $\cal Z$ for the junction-cavity system
\begin{equation}
{\cal Z}=\int {\cal
D}[\varphi] {\cal D}[q_r] \exp\{-S[\varphi , q_r]\}
\label{partition}
\end{equation}
where $S[\varphi, q_r]=\int d\tau {\cal L}$ is the action and $\cal L$ is the Lagrangian
of Eq. (\ref{Lagran}).   By noting that shape distortion of the fluxon due to weak
perturbation (i.e., small $\cal F$) is negligible, we may rewrite the partition function
$\cal Z$ in terms of $q(\tau)$ and $q_r(\tau)$ as
\begin{equation}
{\cal Z} = \int {\cal D}[q] \int {\cal D}[q_r] e^{-S[q, q_r]}~.
\end{equation}
Also by noting that the Lagrangian  ${\cal L}_{coup}$ of Eq. (\ref{coup}) which accounts
for the interaction between LJJ and resonant cavity is linear in both coordinates $q_r$
and $\varphi$, we separate the partition function $\cal Z$ into the resonant cavity and
fluxon contribution by expressing ${\cal Z}={\cal Z}_{res} {\cal Z}_{fluxon}$.   The
resonant cavity (${\cal Z}_{res}$) and fluxon (${\cal Z}_{fluxon}$) contribution to
$\cal Z$ are given, respectively, as ${\cal Z}_{res}=\int {\cal D}[q_r(\omega_n)]\exp
\{-S_{res} [q_r(\omega_n)]\}$ and ${\cal Z}_{fluxon}=\int {\cal D}[q(\tau)]\exp
\{-S_{eff} [q(\tau)]\}$.  The action for the resonant cavity contribution $S_{res}[q_r]$
is given by
\begin{eqnarray}
S_{res} [q_r]=T \sum_{\omega_n} {M_{osc} \over 2}
\left( \omega_n^2 + \omega_r^2 \right) \times~~~~~~~~~~~~
\nonumber \\
\left[ q_{r,n} + {2\pi g_E q_n \omega_n^2 \over
M_{osc}(\omega_n^2 + \omega_r^2)}
\right] \left[ q_{r,-n} + {2\pi g_E q_{-n} \omega_n^2 \over
M_{osc}(\omega_n^2 + \omega_r^2)} \right]
\end{eqnarray}
where $q_{r,n}=q_r(\omega_n)$, $q_n=q(\omega_n)$, $\omega_n=2\pi n T$ is the Matsubara
frequency, and $T$ is the temperature.  The action for the fluxon contribution
$S_{eff} [q]$ is given by
\begin{eqnarray}
S_{eff} [q] = \int d\tau \left[ {M_e \over 2}\sum_{i=1}^2 {\dot q}_i^2 +
V(q) + {{\bar g}_E^2 \omega_r^2 \over {1-{\cal S}^2}} (\sum_{i=1}^2 q_i )^2 \right]
\nonumber \\
- {2{\bar g}_E^2 \over {1-{\cal S}^2}} \int d\tau~
{\dot q}_1 {\dot q}_2 ~~~~~~~~~~~~~~~~~~~~~~~
\nonumber \\
- {{\bar g}_E^2 \over {1-{\cal S}^2}} \int d\tau d\tau'
K(\tau-\tau')
\sum_{i=1}^2 q_i(\tau) \sum_{i=1}^2 q_i(\tau')~~~~
\label{actioneff}
\end{eqnarray}
where  ${\dot q}_i = dq_i/d\tau$, $M_e$ denotes the renormalized fluxon mass
\begin{equation}
M_{e}=M \left( 1 - {1 \over M}
{2{\bar g}_E^2 \over {1-{\cal S}}^2} \right)
\end{equation}
due to the spatially uniform junction-cavity interaction and the $M$ denotes the rest mass
of the fluxon.  The mass $M_e$ accounts for the renormalization effect of both
junction-cavity and magnetic induction interaction.  The bare potential $V(q)=
V(q_1, q_2)$ is given\cite{GW} by
\begin{equation}
V(q)= -\sum_{i=1}^2 \left ( 2\pi f_i q_i +
{2\epsilon_i \over \cosh^2 q_i} \right ) -
{8{\cal S} (q_1 -q_2 ) \over \sinh (q_1 - q_2 )}~.
\label{poten}
\end{equation}
Here, the fluxon potential $V(q)$ includes the effects from the three contributions: (i)
the potential tilting effect ($f$), (ii) the pinning effect ($\epsilon$), and (iii) the
magnetic induction effect ($\cal S$).  The third term in [~] of Eq. (\ref{actioneff})
accounts for the potential renormalization due to junction-cavity interaction.  This
renormaliztion is similar to that for the electronic tunneling process with phonon
coupling.\cite{Se}  In the discussion below, we refer ${\bar g}_E^2=2\pi^2g_E^2/M_{osc}$
as the strength of junction-cavity interaction.  The cavity kernel $K(\tau - \tau')$ in
the third term of Eq. (\ref{actioneff}) is given by
\begin{equation}
K(\tau)={\omega_r^3 \over 2}~
{\cosh(\omega_r/2T - \omega_r \vert \tau \vert ) \over
\sinh(\omega_r/2T)} ~.
\label{kernelo}
\end{equation}
at non-zero temperature $T$. This term accounts for the non-local effect arising from the
junction-cavity interaction.

After the calcultion , the oscillator coordinate $q_r$ in the partition function  $\cal Z$
of Eq. (\ref{partition}) is decoupled from the center coordinate $q$.  This separation
allows us to integrate out the $q_r$-coordinate.   Hence, in discussions below, we
consider the fluxon contribution ${\cal Z}_{fluxon}$ to the partition function which is
described by the action $S_{eff}$.  Using $S_{eff}$, we discuss how the junction-cavity
interaction affects both one-fluxon and two-fluxon tunneling in LJJs.

\section{macroscopic quantum tunneling in single junction}

\begin{figure}[t]
\includegraphics[width=6.3cm]{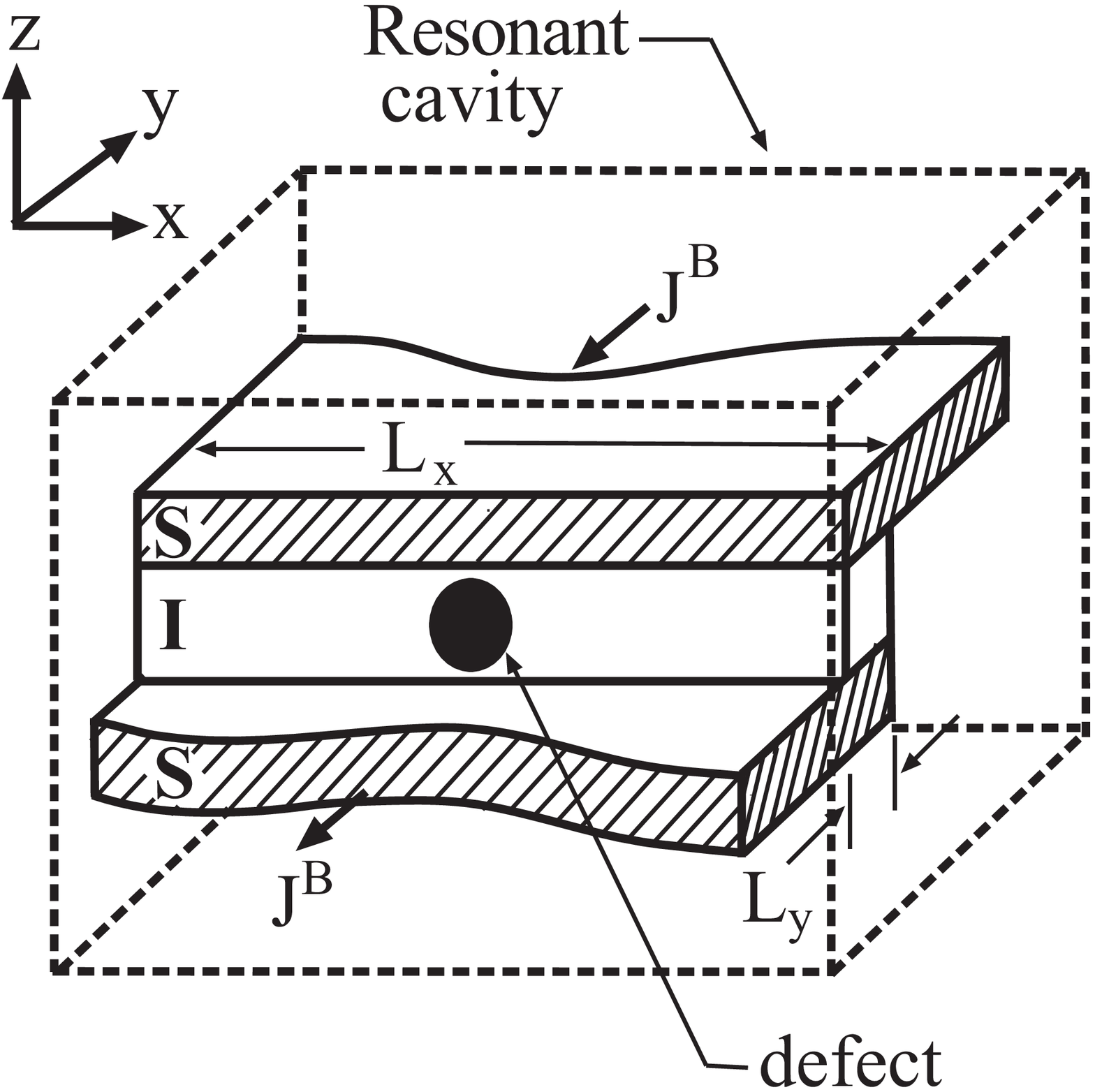}
\caption{ A LJJ is shown schematically as an insulator ($I$) layer
is sandwiched between two superconductor ($S$) layers. $L_x$ and
$L_y$ denote the dimensions in $x-$ and $y-$direction,
respectively. $J^B$ denotes the bias current density.  The filled
circle represents microresistor (i.e., pinning center), and the
dashed box represents resonant cavity.}
\label{ljj}
\end{figure}

\begin{figure}[t]
\includegraphics[width=5.5cm]{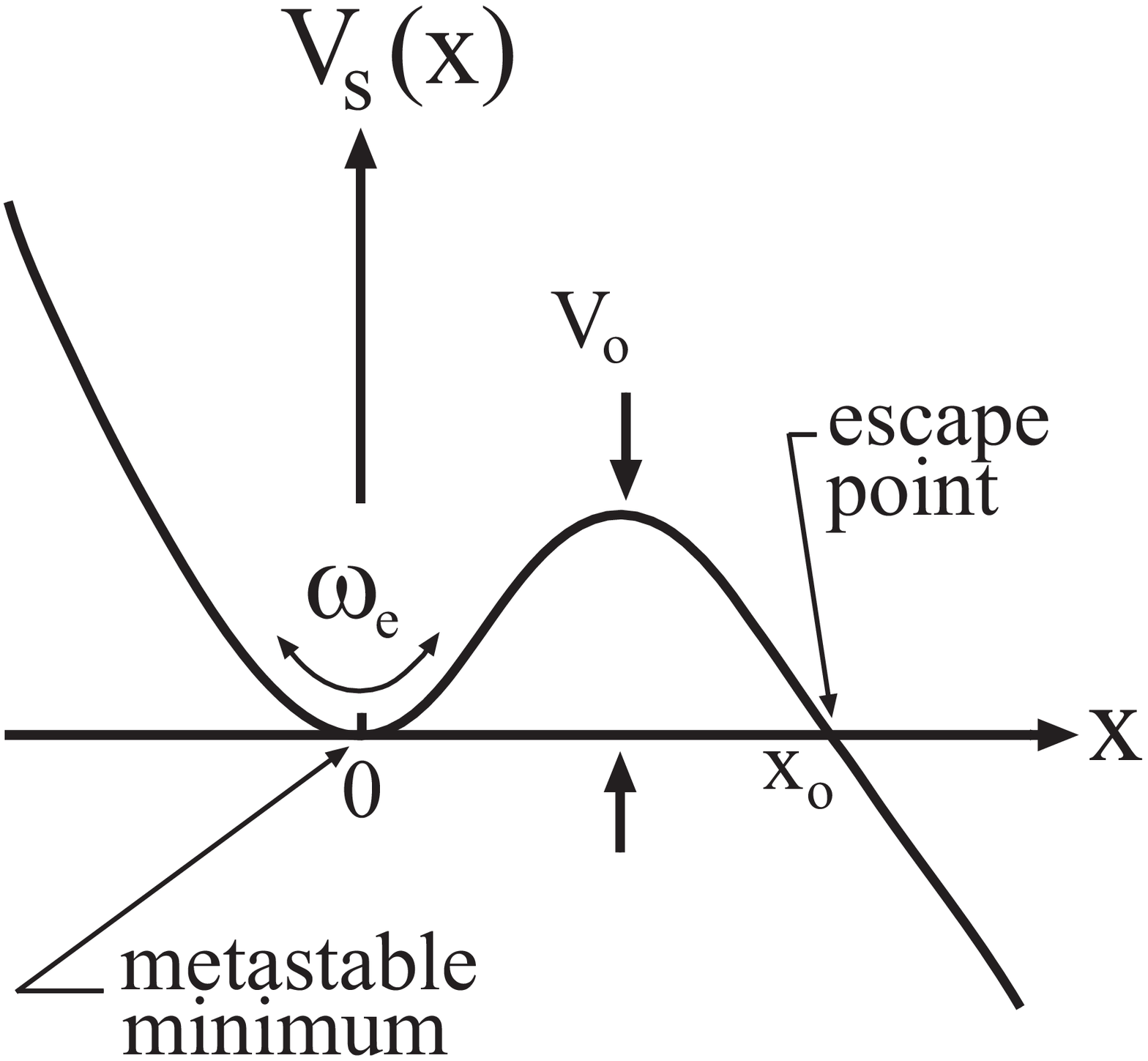}
\caption{The fluxon potential $V_s$ due to both the bias current density
and microresistor in a single LJJ is schematically illustrated.}
\label{potenfig}
\end{figure}

We now examine the effects of resonant cavity on MQT from the metastable state in a single
LJJ obtained by implanting a microresistor in the insulator layer and by applying the
bias current ($J^B$) as shown in Fig. \ref{ljj}.  The dimensions of the junction,
compared to the Josephson length $\lambda_J$, are chosen so that $L_x \gg \lambda_J$ and
$L_y \ll \lambda_J$.  These choices are made to enhance the quantum effect at low
temperatures.  We describe MQT of the fluxon by starting with the action $S_{eff}^s [q]$
for the LJJ given by
\begin{eqnarray}
S_{eff}^s [q] &=& \int d\tau \left [ {M_{e} \over 2} {\dot q}^2
+ V_s(q) + {\bar g}_E^2 \omega_r^2 q^2 \right ]
\nonumber \\
&-& {\bar g}_E^2 \int d\tau d\tau' K(\tau - \tau')~
q(\tau)q(\tau').
\label{actions}
\end{eqnarray}
Here, the action $S_{eff}^s [q]$ is obtained from $S_{eff}[q]$ of Eq. (\ref{actioneff}),
by setting ${\cal S}=0$ (i.e., no magnetic induction effect), $q_1=q$, and $q_2=0$.
Following Caldeira and Leggett, we may simplify $S_{eff}^s [q]$ by making a usual
substitution of $q(\tau)q(\tau')=[q^2(\tau) + q^2(\tau')]/2 - [q(\tau)-q(\tau')]^2/2$.
We note that the first two terms of this substitution cancel the potential renormalization
contribution (i.e., ${\bar g}_E^2 \omega_r^2 q^2$ term) arising from the junction-cavity
interaction.  With this cancellation, the action $S_{eff}^s [q]$ becomes similar to that
for the dissipative system,\cite{CL} but the fluxon mass is now renormalized to
\begin{equation}
M_{e}=M \left( 1 - {2{\bar g}_E^2 \over M} \right)
\end{equation}
and $\beta$ is replaced by the junction-cavity interaction strength (i.e., $\beta
\rightarrow {\bar g}_E^2$).  The renormalized mass $M_e$ accounts for the effects of the
uniform $\bf E$ field in the cavity.  The bare fluxon potential $V_s (q)$ is
given by
\begin{equation}
V_s(q)= -2\pi f q - {2\epsilon \over \cosh^2 q} .
\label{spot}
\end{equation}
Here the bias current density $f=f_c-\delta_f$ is measured in terms of the deviation
$\delta_f$ from the critical value $f_c=4\epsilon/(3\sqrt{3} \pi)$.   The potential
$V_s(q)$ may be approximated by a quadratic-cubic potential as shown schematically in
Fig. \ref{potenfig}.  The cavity kernel $K(\tau - \tau')$  of Eq. (\ref{kernelo})
describing the non-local effect due to the junction-cavity interaction simplifies to
\begin{equation}
K(\tau - \tau') = {\omega_r^3 \over 2}
e^{-\omega_r \vert \tau - \tau' \vert}~
\label{kerqubit}
\end{equation}
in the $T=0$ limit.

The action $S_{eff}^s [q]$ of Eq. (\ref{actions}) indicates that the resonant cavity
yields i) fluxon mass renormalization and ii) non-local effects.  The mass
renormalization modifies the oscillation frequency about the metastable point, as shown
in Fig. \ref{potenfig}.  This change may be easily seen by computing the oscillation
frequency $\omega_e$ at the metastatble state (i.e., local minimum) as
\begin{equation}
\omega_{e}=\left[ {1 \over M_{e}}{d^2{\bar V}_s(0) \over dx^2} \right]^{1/2}
\approx\omega_o \left(
1 + {{\bar g}_E^2 \over M} \right) ,
\label{freqrenormal}
\end{equation}
where $\omega_o$ is the oscillation frequency at the metastable point in the absence of
the resonant cavity.  The non-local contribution due to junction-cavity interaction is
similar to that for the dissipative system, but to determine the size of this contribution
more calculation is needed.

To estimate the size of these two contributions from the junction-cavity interaction, we
compute the MQT rate\cite{CL,Lang} given by
\begin{equation}
\Gamma_{cav} (0) = {\cal A}_{cav}(0) e^{-{\cal B}_{cav}(0)}
\label{tunrate}
\end{equation}
at $T=0$.  Here, the prefactor ${\cal A}_{cav} (0)$ is given by
\begin{equation}
{\cal A}_{cav}(0) =\sqrt{60}~ \omega_e \left(
{{\cal B}_{o,cav} \over 2\pi} \right)^{1/2}
\end{equation}
and the bounce exponent ${\cal B}_{cav}(0)= {\cal B}_{o, cav} + \delta{\cal B}_{cav}$
includes both the local contribution ${\cal B}_{o, cav}$ of
\begin{equation}
{\cal B}_{o, cav}= \int_{-\infty}^{\infty} d\tau
\left [ {M_{e} \over 2} {\dot q}^2
+ V_s(q) \right ]
\label{tunnelexp}
\end{equation}
and the non-local contribution $\delta{\cal B}_{cav}$ of
\begin{equation}
\delta{\cal B}_{cav} = {\bar g}_E^2
\int_{-\infty}^{\infty} d\tau \int_{-\infty}^{\infty} d\tau'
K(\tau - \tau') \left [ q(\tau) -q(\tau') \right ]^2~.
\label{non-local}
\end{equation}
These two contributions, ${\cal B}_{o, cav}$ and $\delta{\cal B}_{cav}$, to
${\cal B}_{cav}(0)$ are evaluated explicitly to estimate their size.

\begin{figure}[t]
\includegraphics[width=6.5cm]{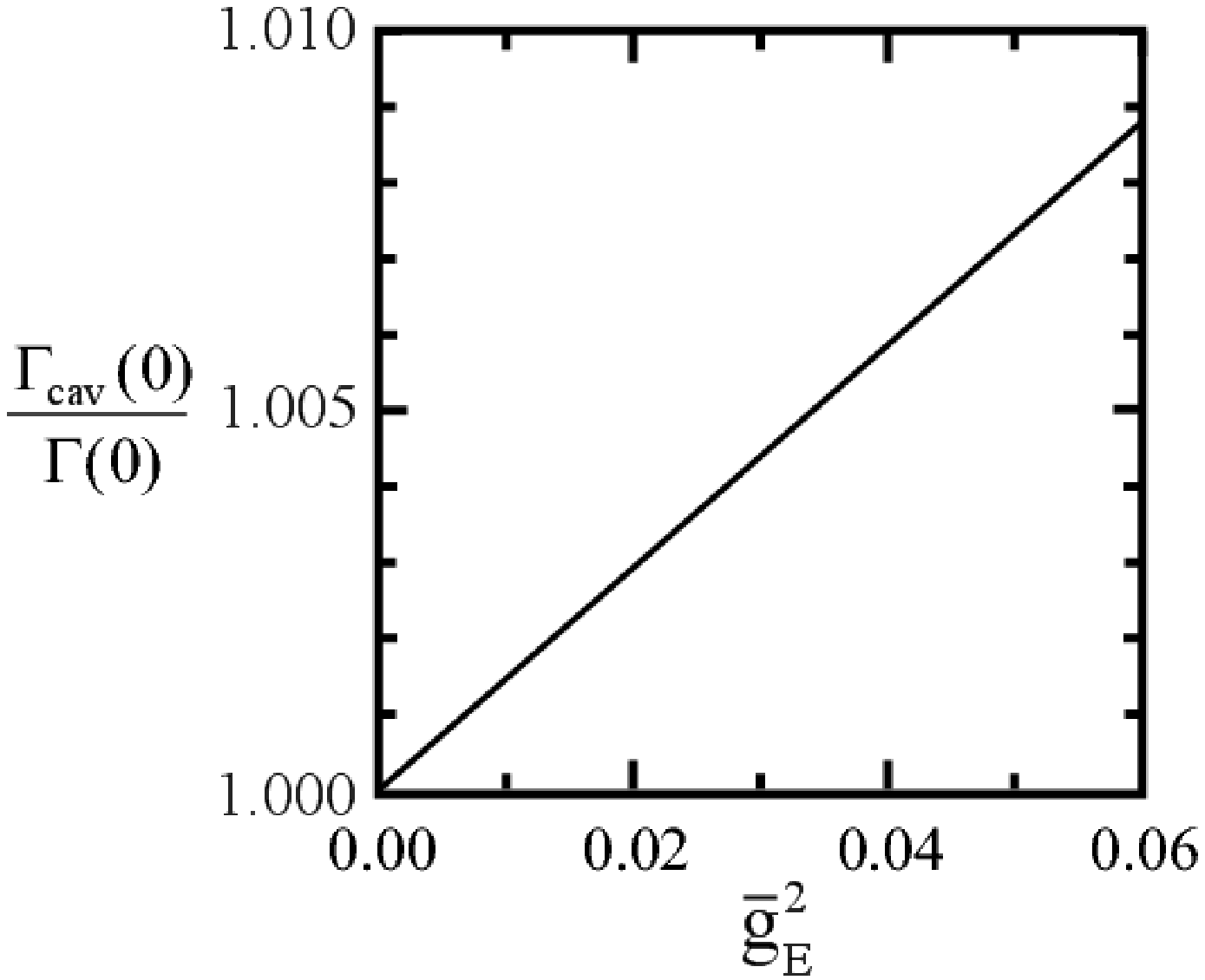}
\caption{The ratio of the tunneling rates $\Gamma_{cav}(0)/\Gamma(0)$
is plotted as a function of the junction-cavity coupling
strength ${\bar g}_E^2$ to illustrate the size of enhancement.}
\label{TunRateSingleImp}
\end{figure}

The local contribution ${\cal B}_{o,cav}$ may be computed easily by approximating $V_s (q)$
of Eq. (\ref{spot}) as a usual quadratic-plus-cubic potential of
\begin{equation}
{\bar V}_s(x) = V_s(q) - V_s(q_o) \approx {27 V_o \over 4}
\left ({\bar x}^2 -{\bar x}^3 \right )
\end{equation}
where ${\bar x}=x/x_o$, $x=q-q_o$, and $V_o=[8\pi^3\delta_f^3/(\sqrt{3}\epsilon)]^{1/2}$
is the barrier potential for the fluxon.  Here $q_o$ is the position of the metastable
point and $x_o=9\sqrt{3}M_{e}\omega_{e}^2/32\epsilon$ is the escape point as shown in
Fig. \ref{potenfig}.  The evaluation of ${\cal B}_{o,cav}$ yields
\begin{equation}
{\cal B}_{o,cav} = 2 \int_o^{x_o} dx
\left[ 2M_{e} {\bar V}_s(x) \right]^{1/2} =
{36 V_o \over 5 \omega_{e}}~.
\end{equation}
Using this result, we estimate the local contribution to enhancement of the tunneling rate
due to the resonant cavity.  The ratio of the MQT rates, $\Gamma_{cav}(0) / \Gamma(0)$, is
given by
\begin{equation}
{\Gamma_{cav}(0) \over \Gamma(0)} \approx 1 +
{{\bar g}_E^2 \over 2M} \left( 1 + {72 \over 5}{V_o \over \omega_o} \right )~,
\label{tunratio}
\end{equation}
where $\Gamma (0)$ is the tunneling rate in the absence of the resonant cavity (i.e.,
${\bar g}_E^2 =0$).  Equation (\ref{tunratio}) indicates that the tunneling rate increases
with increasing junction-cavity interaction strength ${\bar g}_E^2$.  In Fig.
\ref{TunRateSingleImp}, we plot the numerically computed ratio $\Gamma_{cav}(0)/\Gamma(0)$
as a function of ${\bar g}_E^2$ to illustrate its enhancement in the weak-coupling regime
(i.e., ${\bar g}_E^2 \ll 1$).  The curve indicates that enhancement of
$\Gamma_{cav}(0)/\Gamma(0)$ is less than 1$\%$.

\begin{figure}[t]
\includegraphics[width=6.3cm]{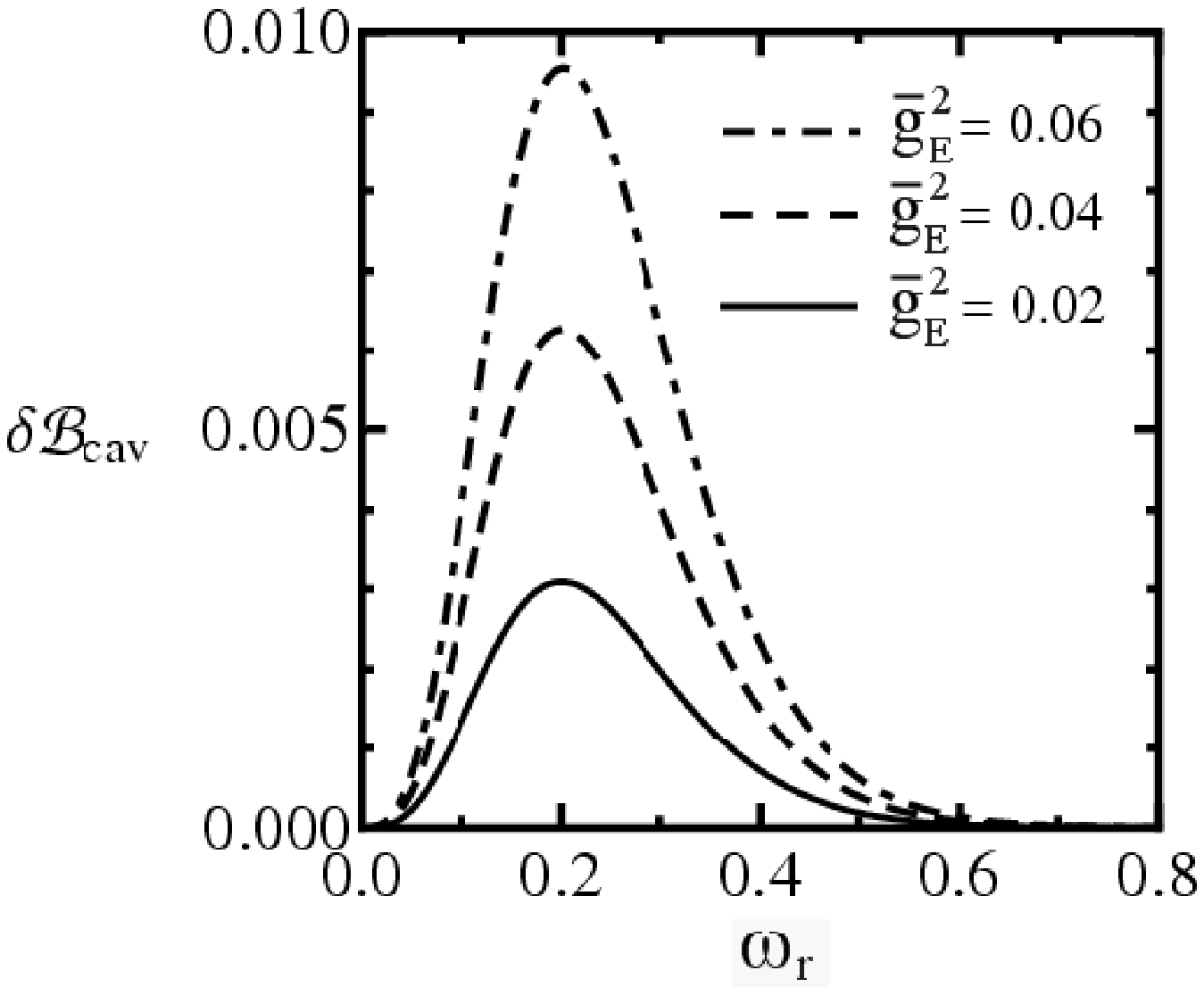}
\caption{The non-local contribution $\delta {\cal B}_{cav}$ to the
bounce exponent  ${\cal B}_{cav}(0)$ is plotted as a function of
$\omega_r$ for ${\bar g}_E^2 = 0.02$ (solid line), 0.04
(dashed line) and 0.06 (dot-dashed line). }
\label{deltaB}
\end{figure}

The non-local contribution $\delta {\cal B}_{cav}$ to ${\cal B}_{cav} (0)$ of Eq.
(\ref{tunrate}) reduces the tunneling rate $\Gamma_{cav}(0)$.  The size of this reduction
is estimated by evaluating $\delta{\cal B}_{cav}$ of Eq. (\ref{non-local}) by writing
\begin{equation}
\delta {\cal B}_{cav} = {{\bar g}_E^2 \omega_r^3 x_o^2 \over
\omega_{e}^2} \int_{-\infty}^{\infty} d{\bar \tau} d{\bar \tau'}
e^{-{2\omega_r \over \omega_e}\vert {\bar\tau} - {\bar\tau'} \vert}
\left [{\bar x}({\bar \tau}) -{\bar x}({\bar \tau'}) \right ]^2~,
\label{non-local1}
\end{equation}
where ${\bar x}(\tau) = {\rm sech}^2(\omega_{e}\tau/2)$.   We note that ${\bar x}(\tau)$
is the solution to the equation of motion for the quadratic-plus-cubic potential in the
absence of the non-local effect.  We evaluate Eq. (\ref{non-local1}) and obtain
\begin{equation}
\delta {\cal B}_{cav} = 2{\bar g}_E^2 \left({9\sqrt{3} M_{e} \over
16\epsilon}\right)^2
{\omega_r^5 \over {\rm sinh}^2(\pi\omega_r/\omega_{e})}~.
\end{equation}
The result for $\delta {\cal B}_{cav}$ indicates that the non-local contribution increases
almost linearly with ${\bar g}_E^2$ in the weak-coupling regime and has a strong
dependence on the frequency $\omega_r$ of the cavity mode.  At low cavity frequencies
($\omega_r \ll 1$), the non-local contribution varies as
$\delta {\cal B}_{cav} \propto \omega_r^3$.  At high cavity frequencies
($\omega_r \gg 1$), on the other hand, it varies as
$\delta {\cal B}_{cav} \propto \omega_r^5 \rm{exp} (-2\pi\omega_r/\omega_{e})$.  To
illustrate the cavity frequency dependence, we plot $\delta {\cal B}_{cav}$ as a function
of $\omega_r$ for ${\bar g}_E^2 =$ 0.02 (solid line), 0.04 (dashed line), and 0.06
(dot-dashed line) in Fig. \ref{deltaB}.  The curves indicate that $\delta {\cal B}_{cav}$
vanishes both in the low and high cavity frequency $\omega_r$ limits.  Hence, the
non-local effects on the tunneling rate $\Gamma_{cav}(0)$ is negligible near these limits.

\section{macroscopic quantum tunneling in coupled junctions}

\begin{figure}[t]
\includegraphics[width=6.3cm]{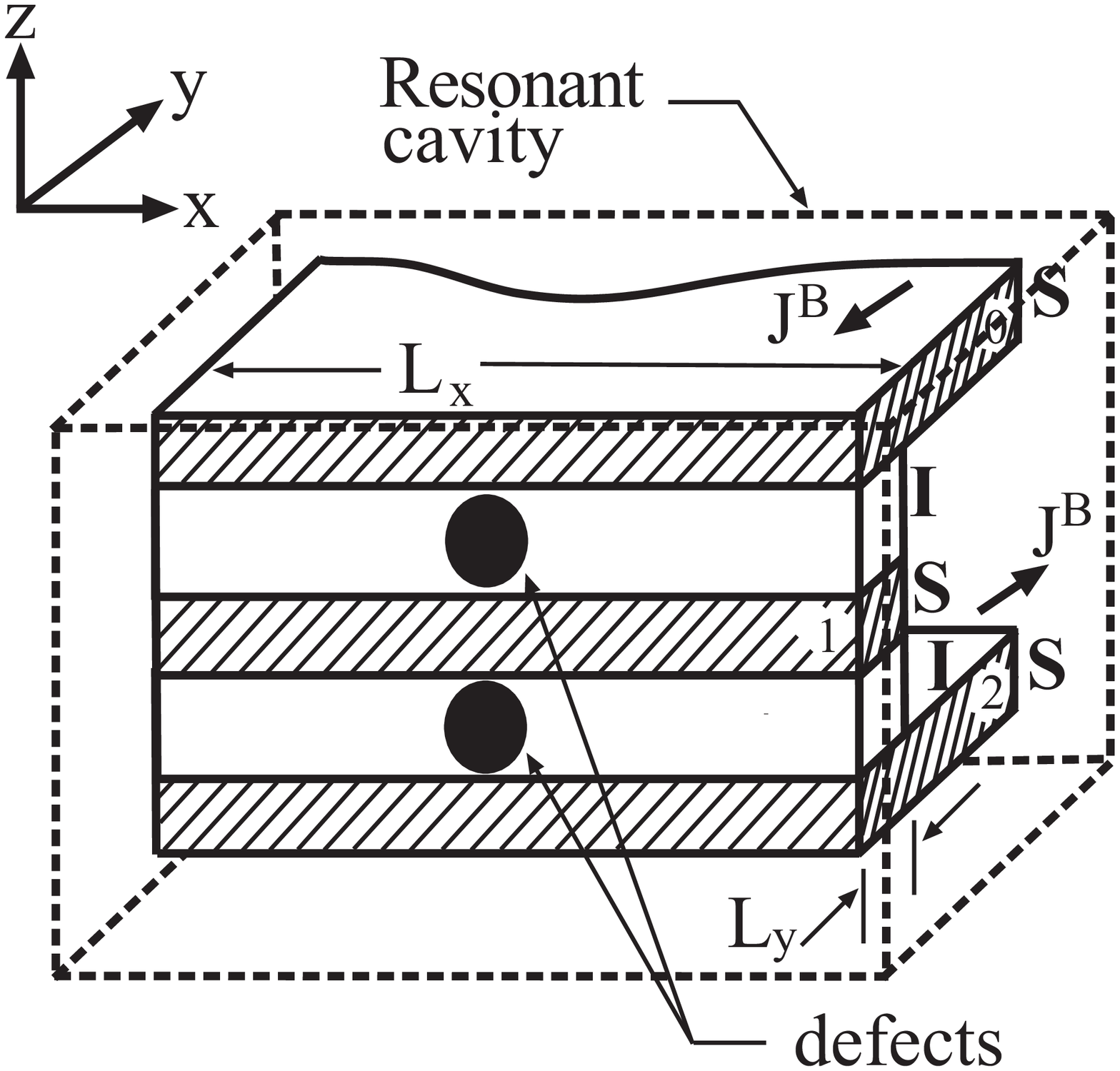}
\caption{Two LJJs with a vertical column of two microresistors
is shown schematically.  $L_x$ and $L_y$ denote the
dimensions in $x-$ and $y-$direction, respectively.  $J^B$
denotes the bias current density.  The filled circles represent
the microresistors.}
\label{ljj2s}
\end{figure}

In this section, we estimate the effects of resonant cavity on the tunneling rate of the
phase-locked fluxons from the metastable state in two coupled LJJs.  Here the fluxons are
trapped by the microresistor on each insulator (I) layer, shown schematically in Fig.
\ref{ljj2s}.  Earlier studies\cite{KM} indicate that uncorrelated one-fluxon tunneling is
the dominant process in the absence of resonant cavity.  However, phase-locking between
the fluxons in two LJJs becomes enhanced in the resonant cavity.  This enhancement may be
seen more easily from the effective action $S_{eff}[q]$ for the two coupled LJJs of Eqs.
(\ref{actioneff}) and (\ref{poten}) written in the rotated coordinates $(q_+, q_-)$ as
\begin{eqnarray}
S_{eff} [q]= \int d\tau \left [ {M_{e} \over 2} {\dot q}_+^2 + {M \over 2} {\dot q}_-^2 +
V({\bf q}) + {{\bar g}_E^2 \omega_r^2 \over 1-{\cal S}^2} q_+^2 \right ]
\nonumber \\
- {2{\bar g}_E^2 \over {1-{\cal S}^2}} \int d\tau d\tau' K(\tau - \tau')~
q_+(\tau) q_+(\tau') ,~~~~
\end{eqnarray}
where $q_{\pm} = (q_1 \pm q_2)/\sqrt{2}$.  The action $S_{eff} [q]$ indicates that the
potential for the in-phase mode, $(q_+, 0)$, is renormalized by the junction-cavity
interaction while the out-of-phase mode, $(0, q_-)$, is not.   Also, the non-local
contribution appears only for the motion in the $q_+$ direction.  The bare fluxon
potential $V({\bf q})= V(q_+, q_-)$ of
\begin{eqnarray}
V(q_+, q_-)= - 2\sqrt{2}\pi f q_+ -
{8 \sqrt{2}{\cal S} q_- \over \sinh \sqrt{2} q_-} ~~~~~~~~~~~~~~
\nonumber \\
- 2\epsilon \left[{1 \over \cosh^2
\left( {{q_+ + q_-} \over \sqrt{2}}  \right) }
+ {1 \over \cosh^2
\left( {{q_+ - q_-} \over \sqrt{2}} \right )} \right]
\end{eqnarray}
for $f_1 = f_2 =f$ and $\epsilon_1 = \epsilon_2 = \epsilon$, indicates that the
one-dimensional potential along the $(q_+,0)$ direction (i.e., $V(q_+,0)$) corresponding
to the in-phase mode becomes identical to $V_s(q)$ of Eq. (\ref{spot}) under the
transformation of $2f \rightarrow f$, $2\epsilon \rightarrow \epsilon$, and
$q_+/\sqrt{2} \rightarrow q$.  This similarity reflects that the phase-locked fluxons
moving in the ($q_+,0$) direction (i.e., $q_1=q_2$) behave as a single fluxon.  However,
the one-dimensional potential for the out-of-phase mode (i.e., $V(0,q_-)$ or along the
$(0,q_-)$ direction) behaves as a potential well near the metastable point
${\bf q}^o = (q^o_+,q^o_-)$, determined from the condition
$[\partial V({\bf q})/\partial q_+]_{q_-}= [\partial V({\bf q})/\partial q_-]_{q_+}=0$.

\begin{figure}[t]
\includegraphics[width=7.0cm]{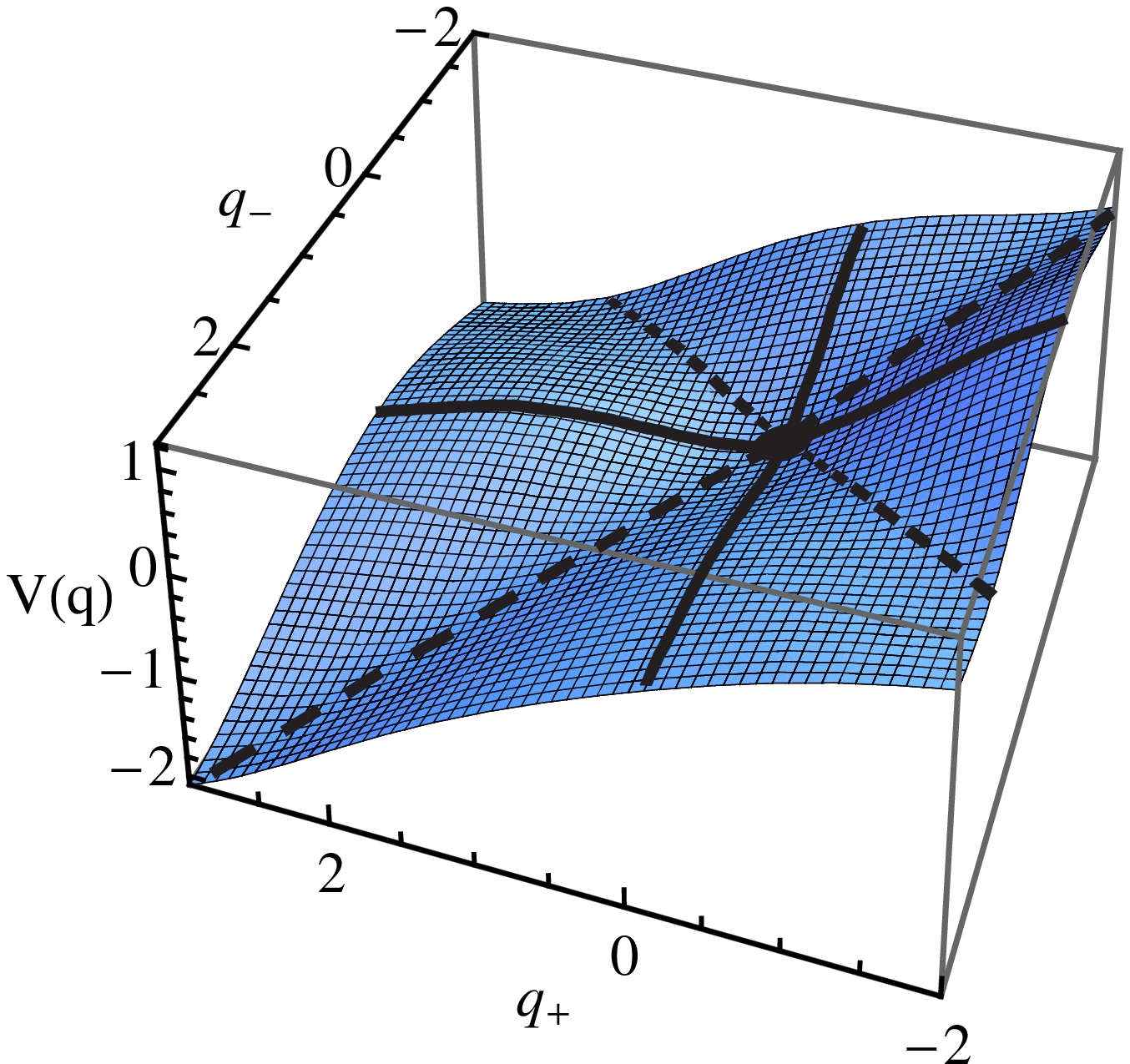}
\caption{The potential $V_Q(q_+, q_-)$ surface is plotted for
$\epsilon =0.269$ and ${\cal S}=-0.05$.   The filled circle
represents the position of the metastable state.  The dashed
and solid lines denote the most probable escape paths (MPEPs)
for one-fluxon and two-fluxon tunneling, respectively.   }
\label{potsurf}
\end{figure}

To illustrate these phase-locking modes, we plot the potential $V(q_+, q_-)$ in Fig.
\ref{potsurf} for $f=0.06$, $\epsilon = 0.269$, and ${\cal S}=-0.05$.  Here, the value for
$\epsilon$ and $\cal S$ are chosen so that when a vertical stack\cite{KOD} of two
interacting JVQs are fabricated using coupled LJJs and microresistors only one quantum
state is bound on each side of the double-well potential.  The metastable point
${\bf q}^o$ is denoted by the solid circle.  The solid lines indicate that the potential
is metastble for the in-phase mode (i.e., along the $(q_+,0)$ direction), but it behaves
as a well for the out-of-phase mode (i.e., along the $(0,q_-)$ direction).  These curves
show that tunneling of the in-phase mode from the metastable state is more favorable than
that for the out-of-phase mode.

The tunneling rate $\Gamma_{cav} (0)$ from ${\bf q}^o$ can be estimated by summing
over the contribution from all paths of escape, but the dominant contribution comes
from the most probable escape path (MPEP) in which $S_{eff}$ is the minimum.\cite{BBW}
For the physical parameters chosen in Fig. \ref{potsurf}, the MPEPs correspond to
one-fluxon tunneling, indicated by the dashed lines.  The MPEPs are determined by the two
competing energies: (i) the pinning energy (${\cal E}_{pin} = \vert E_{pin}\vert$) and
(ii) the magnetic induction interaction energy (${\cal E}_{int}=\vert E_{int} \vert$).
When ${\cal E}_{int} \gg {\cal E}_{pin}$, the fluxons are not pinned at the microresistor
sites but maintain a large separation distance.\cite{GW}  However, when
${\cal E}_{int} \ll {\cal E}_{pin}$, the one-fluxon tunneling processes are favored over
the two-fluxon tunneling processes.

We now estimate the two-fluxon tunneling rate for the in-phase mode.  We simplify the
calculation by using the simialrity between the tunneling of the in-phase mode and the
one-fluxon tunneling process discussed in Sec. III.  When the bias current $f$ is less
than the critical value $f_c$ (i.e., $f=f_c- \delta_f$ with
$0<\delta_f\ll f_c = 4\epsilon/(3\sqrt{3} \pi)$, the potential along the path $(q_+, 0)$
has the metastable state, as illustrated in Fig. \ref{potenfig}.  The potential $V(q_+,0)$
may be approximated as the quadratic-plus-cubic form of
\begin{equation}
V({\bar q}_+, 0) \approx {27 V_o^{t} \over 4}
({\bar q}_+^2 - {\bar q}_+^3)~,
\end{equation}
where ${\bar q}_+ = (q_+ - q_+^o)/q_+^e$, $q_+^e$ is the escape point and
$V_{o}^{t}=2[d^2V(q_+^o,0)/dq_+^2]^3/3 [d^3V(q_+^o,0)/dq_+^3]^2$ denotes the potential
barrier height for two-fluxon tunneling.   We note that $q_+^e$ is similar to $x_o$ in
Fig. \ref{potenfig}.  Also, similar to the single LJJ, the semiclassically estimated
two-fluxon tunneling rate of
$\Gamma_{cav}^{t} (0)={\cal A}_{cav}^{t} \exp[{\cal B}_{cav}^{t}(0)]$ at $T=0$ depends on
both the barrier height and oscillation frequency.   The factor ${\cal A}_{cav}^{t} (0)$
and bounce exponent ${\cal B}_{cav}^{t}(0)$ are calculated in the same way as in Sec. III.
The factor ${\cal A}_{cav}^{t} (0)$ is given by
\begin{equation}
{\cal A}_{cav}^{t}(0) \approx \sqrt{60}~\omega_{e} \left(
{{\cal B}_{o,cav}^{t} \over 2\pi} \right)^{1/2}~.
\label{prefa}
\end{equation}
The local and non-local contributions to the bounce exponents ${\cal B}_{cav}^{t} (0) =
{\cal B}_{o, cav}^{t} + \delta {\cal B}_{cav}^{t}$ are given by
\begin{equation}
{\cal B}_{o, cav}^{t} =2 \int_0^{q_+^e - q_+^o} dq_+
\sqrt{2M_e V(q_+, 0)} \approx {36 V_{o}^{t} \over 5\omega_{e}} ,
\end{equation}
and
\begin{equation}
\delta {\cal B}_{cav}^{t} \approx {2{\bar g}_E^2 \over 1 - {\cal S}^2}
\left({9\sqrt{3} M_{e} \over 16\epsilon}\right)^2 {\omega_r^5 \over
{\rm sinh}^2(\pi\omega_r/\omega_{e})}~,
\end{equation}
respectively.  The result indicates that the two-fluxon tunneling rate
$\Gamma_{cav}^{t} (0)$ in the cavity is enhanced from that $\Gamma^{t}(0)$ in its absence.
Neglecting the non-local contribution, we may write the ratio
$\Gamma_{cav}^{t} (0)/\Gamma^{t}(0)$ as
\begin{equation}
{\Gamma_{cav}^{t} (0) \over \Gamma^{t}(0)} \approx
1+ {{\bar g}_E^2 \over 2M(1-{\cal S}^2)}
\left( 1 + {72 \over 5} {V_o^{t} \over \omega_o} \right)~.
\end{equation}
This enhancement is similar to the tunneling process discussed in Sec. III.  The estimated
value of $\Gamma^{t} (0)$ for the Nb-Al$_2$O$_x$-Nb-Al$_2$O$_x$-Nb junction is
$8.5 \times 10^9$ $s^{-1}$.  This value is obtained by using the experimental
value\cite{Sakai,Kl} of $J_c\sim $2$\times 10^6$A/m$^2$, $\lambda_L\sim$90nm,
$\lambda_J \sim$25$\mu$m, and $\omega_p\sim$90GHz.  Also, we chose $L_y \sim$0.2$\mu$m to
enhance the quantum effect and used the experimentally accessible value\cite{KI} of
$\epsilon=0.269$, ${\cal S}=-0.05$ and $\delta_f\sim$5$\times$10$^{-4}$.  On the other
hand, the potential $V(q_+, q_-)$ along the $(q_+,0)$ direction indicates that the
two-fluxon tunneling rate $\Gamma_{cav}^{t} (0)$ is suppressed from the one-fluxon
tunneling rate $\Gamma_{cav}^{o} (0)$ along either the $q_+ = q_-$ or $q_+ = -q_-$
direction.  This reduction in the tunneling rate is given by
\begin{eqnarray}
{\Gamma_{cav}^{t} (0) \over \Gamma_{cav}^o(0)} \approx
&&\alpha_o \sqrt{V_o^{t} \over V_o^{o}}
e^{-{36(V_o^{t} - \alpha_o^2 V_o^{o}) \over 5 \omega_o}} \times
\nonumber \\
&&\left[ 1 + {36{\bar g}_E^2 (V_o^{t} - \alpha_o^2 V_o^{o}) \over
5\omega_o M(1-{\cal S}^2)} \right]~,
\label{tunto}
\end{eqnarray}
where $\alpha_o = \{[d^2V(q_+^o,0)/dq^2]/[d^2V(q^o,0)/dq^2]\}^{1/4}$ is a constant of order
unity, $V_{o}^{o}=2[d^2V(q^o,0)/dq^2]^3/3[d^3V(q^o,0)/dq^3]^2$ is the one-fluxon
tunneling potential barrier height, $V(q,0)$ is the fluxon potential of Eq. (\ref{poten})
along the $q_+=q_-$ direction, and $q^o$ denotes the position of the metastable point for
one-fluxon tunneling, given by the condition that $dV(q^o,0)/dq=0$.  The ratio
$\Gamma_{cav}^{t}(0)/\Gamma_{cav}^o(0) \ll 1$ for the potential surface in Fig.
\ref{potsurf} reflects that $V_o^{t} \gg V_o^{o}$.

\section{Josephson vortex qubit in resonant cavity}

\begin{figure}[t]
\includegraphics[width=6.3cm]{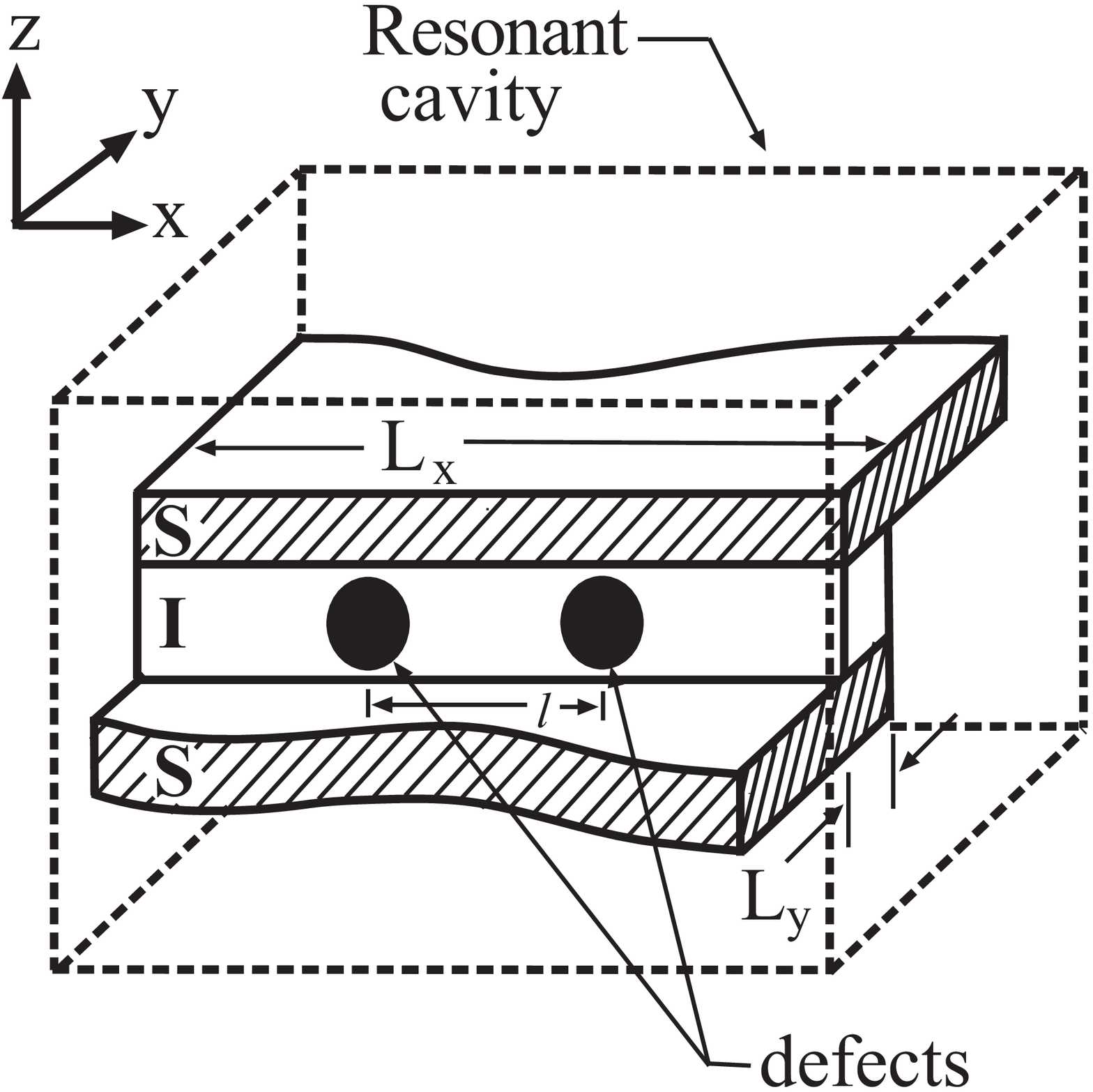}
\caption{A LJJ with two microresistors, representing a Josephson vortex qubit, in a
resonant cavity is shown schematically.  The separation distance between the
microresistors is denoted by $\ell$.  The filled circles and dashed box represent the
microresistors and resonant cavity, respectively. }
\label{jvq}
\end{figure}

We now examine the effects of high-$Q_c$ resonant cavity on JVQ.  The JVQ may be
fabricated by using two closely implanted microresistors in the insulator layer of the
linear LJJ as shown in Fig. \ref{jvq}.  As earlier studies\cite{Cl,KDP,oJVQ} indicate, MQT
of fluxon between the spatially separated minima of double-well potential leads to
splitting of the degenerate ground-state energy.\cite{GWH,Garg}  In this section, we
estimate the effects of junction-cavity interaction on this energy splitting.

The interaction between the LJJ and resonant cavity yields i) fluxon potential
renormalization and ii) non-local contribution to the action.  The effects of these
contributions on the energy splitting may be estimated by starting with the action
$S_{eff}^{Q}$ for the JVQ given by
\begin{eqnarray}
S_{eff}^{Q} [q] &=& \int d\tau \left [ {M_{e} \over 2} {\dot q}^2
+ V_{Q}(q) \right ]~
\nonumber \\
&-& 2{\bar g}_E^2 \int d\tau d\tau' ~K(\tau - \tau')~
q(\tau) q(\tau')~.
\label{actionQ}
\end{eqnarray}
Without loss of generality, we obtain the potential function $V_Q(q)$ from the
double-well potential $V(q)$ of
\begin{equation}
V(q)= {\bar g}_E^2 \omega_r^2 q^2
- {2 \epsilon \over \cosh^2 \left( q-{\ell \over 2} \right )}
- {2 \epsilon \over \cosh^2 \left( q+{\ell \over 2} \right )} ,
\label{Qpot}
\end{equation}
where $\ell$ denotes the separation distance between the two microresistors.  Here, we
have added a constant energy $E_Q$ term to $V(q)$ (i.e., $V_Q(q) = V(q) + E_Q$) so that
$V_Q(q)$ vanishes at the potenital minima.   Here, the potential $V_Q(q)$ may be
characterized by the position of the two minima and the potential barrier height.  In the
discussion below, we do not make the usual substitution of $q(\tau)q(\tau')=[q^2(\tau) + q^2(\tau')]/2 - [q(\tau)-q(\tau')]^2/2$ used in Sec. III.   This approach allows us to
elucidate the origin of the changes in the energy splitting due to the junction-cavity interaction.

\begin{figure}[t]
\includegraphics[width=7.5cm]{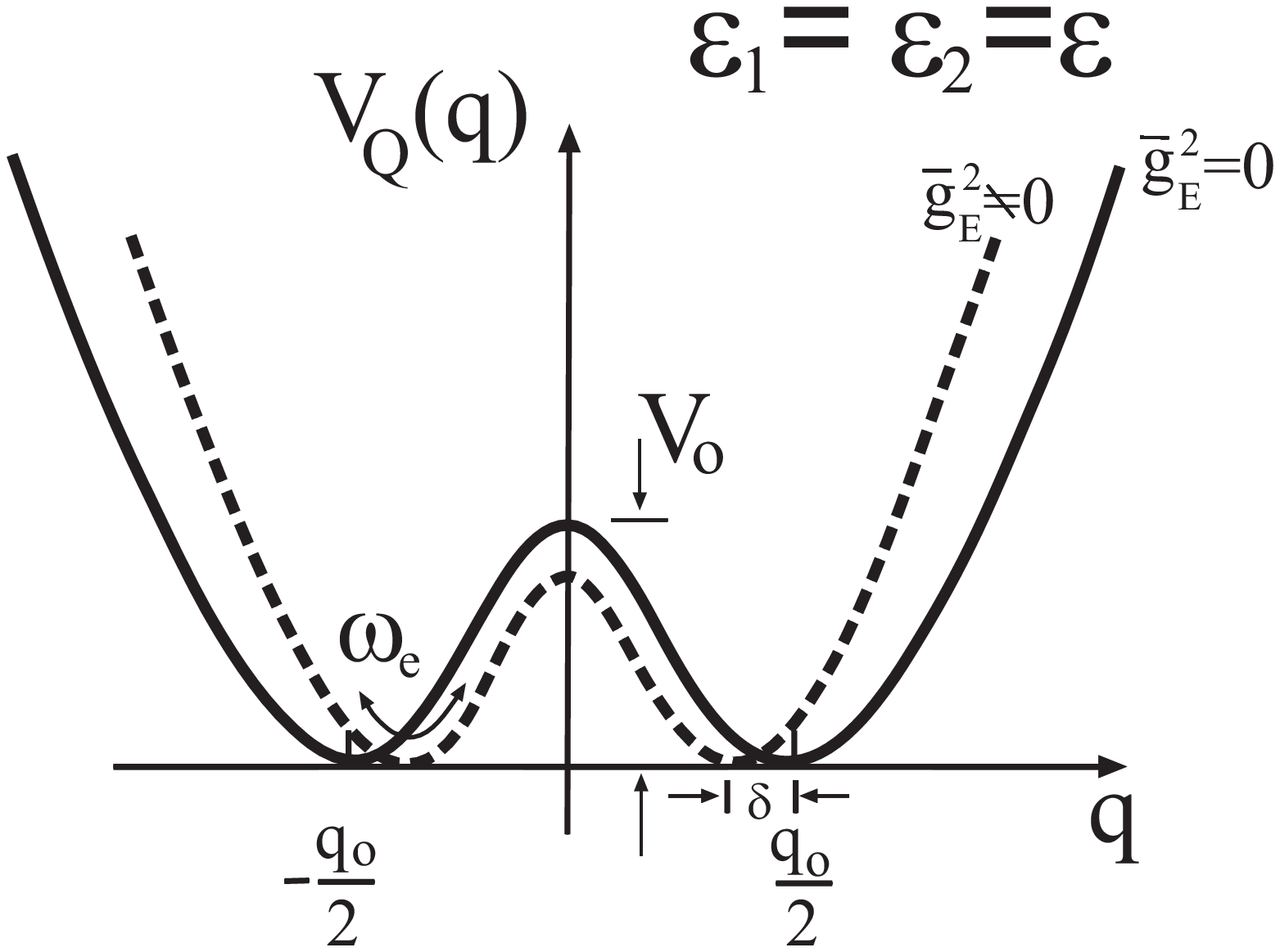}
\caption{A schematic diagram of a double-well potential $V_Q(q)$ due to the two
microresistors in the insulator layer of the LJJ is shown to illustrate the
renormalization of $V_Q(q)$.  The solid and dashed lines represent the potential $V_Q(q)$
in the absence and in the presence of the resonant cavity, respectively.}
\label{doublewellcavity}
\end{figure}

In the absence of the resonant cavity (i.e., ${\bar g}_E^2=0$), the double-well structure
for $V_Q(q)$ with the separation distance $\ell > \ell_o \approx 1.317$ is shown
schematically in Fig. \ref{doublewellcavity} as the solid line.  The two potential minima
are located at $q=\pm q_o/2$ where $q_o$ is determined from
\begin{equation}
\cosh q_o = {\sinh^2 \ell - 1 \over  \cosh \ell}~.
\end{equation}
The energy shift $E_Q$, representing a constant of motion, is given by
\begin{equation}
E_Q = -2\epsilon {\cosh^2 \ell \over \cosh^2 \ell -1}~.
\end{equation}
Also, the potential barrier height $V_o$ between the two minima (i.e., $q=\pm q_o/2$) is
given by
\begin{equation}
V_o = 2 \epsilon \left(
{\cosh \ell - 2 \over \sinh \ell} \right)^2~.
\end{equation}
We note that these quantities change in the resonant cavity, as shown schematically by
the dashed line in Fig. \ref{doublewellcavity}.

In the resonant cavity (i.e., ${\bar g}_E^2 \neq 0$), on the other hand, the JVQ potential
$V_Q(q)$ acquires an additional ${\bar g}_E^2 \omega_r^2 q^2$ term in Eq. (\ref{Qpot}).
This term arises from the coupling between the oscillator coordinate $q_r$ and the center
coordinate $q$ in the coupling Lagrangian ${\cal L}_{coup}$ of Eq. (\ref{coup}) and
accounts for potential renormalization.  The main renormalization effects are the
following: i) the barrier potential height is reduced, ii) the position of the potential
minima become closer together, and iii) the oscillation frequency at the potential minima
is modified.  These effects become amplified with increasing junction-cavity interaction
strength (${\bar g}_E^2$) and resonant frequency ($\omega_r$).

\begin{figure}[t]
\includegraphics[width=6.5cm]{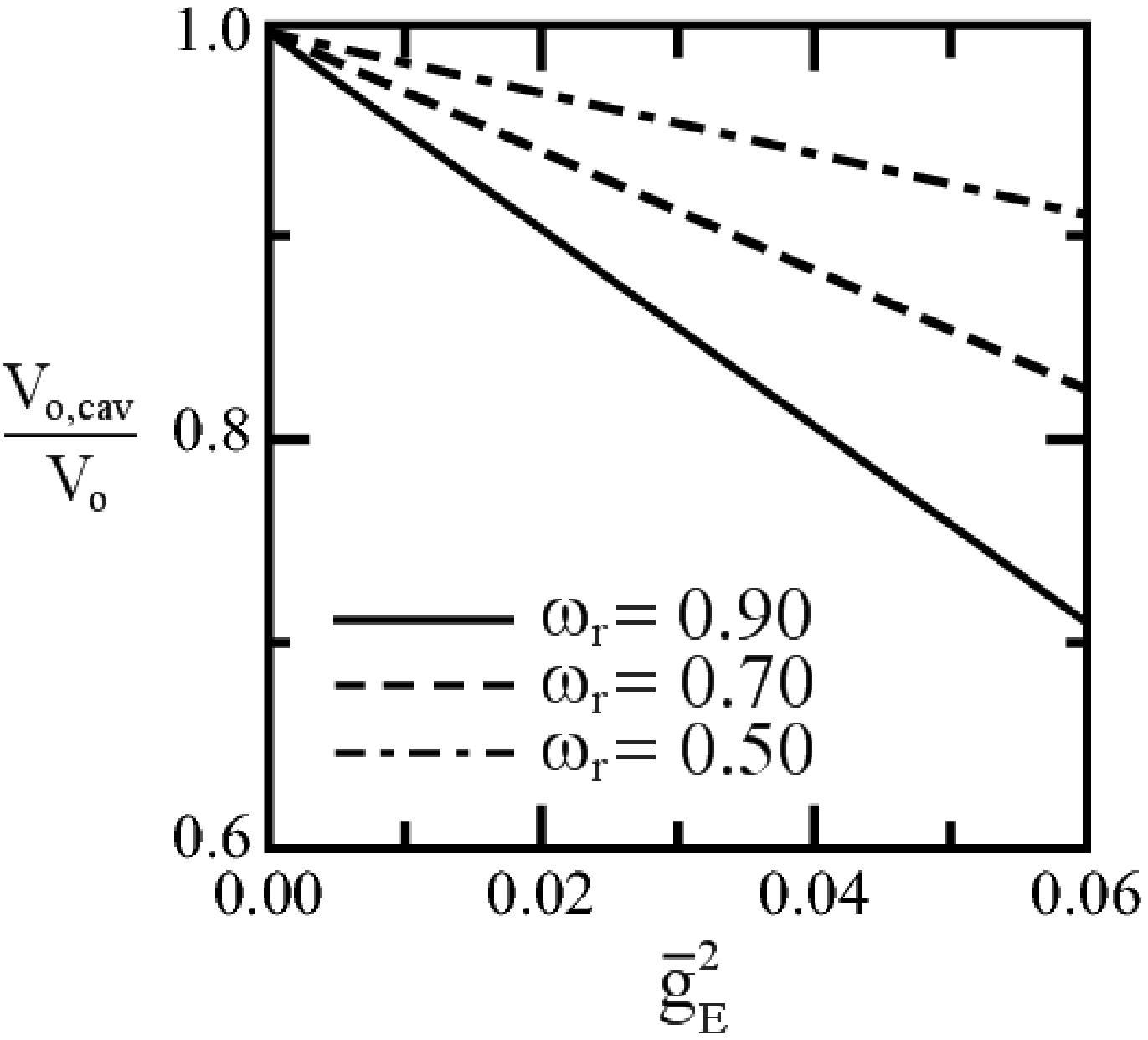}
\caption{The ratio of the potential barrier height $V_{o,cav}/V_o$ is plotted as a
function of the junction-cavity coupling strength ${\bar g}^2_E$ for $\omega_r = 0.50$
(dot-dashed line), 0.70 (dashed line), and 0.90 (solid line) to illustrate the suppression
in the cavity. }
\label{VcavvsV}
\end{figure}

The effects of the junction-cavity interaction on the potential barrier height $V_{o,cav}$
may be estimated straightforwardly.  In Fig. \ref{VcavvsV}, we plot the numerically
computed ratio $V_{o,cav}/V_o$ as a function of ${\bar g}_E^2$ to illustrate the
dependence on the junction-cavity interaction.  The curves for $\omega_r=0.50$
(dot-dashed line), 0.70 (dashed line) and 0.90 (solid line) indicate that the barrier
potential height decreases with increasing ${\bar g}_E^2$ and $\omega_r$.  Also, the
curves indicate that the ratio decreases linearly in the weak coupling regime.  To leading
order in ${\bar g}_E^2$, the potential barrier height $V_{o,cav}$ estimated from the
renormalized potential $V(q)$ of Eq. (\ref{Qpot}) is given by
\begin{equation}
V_{o,cav} \cong V_o -{\bar g}_E^2 \omega_r^2 q_o^2~.
\end{equation}
This decrease in the potential barrier height leads to the increase in the ground-state
energy splitting.

\begin{figure}[t]
\includegraphics[width=6.3cm]{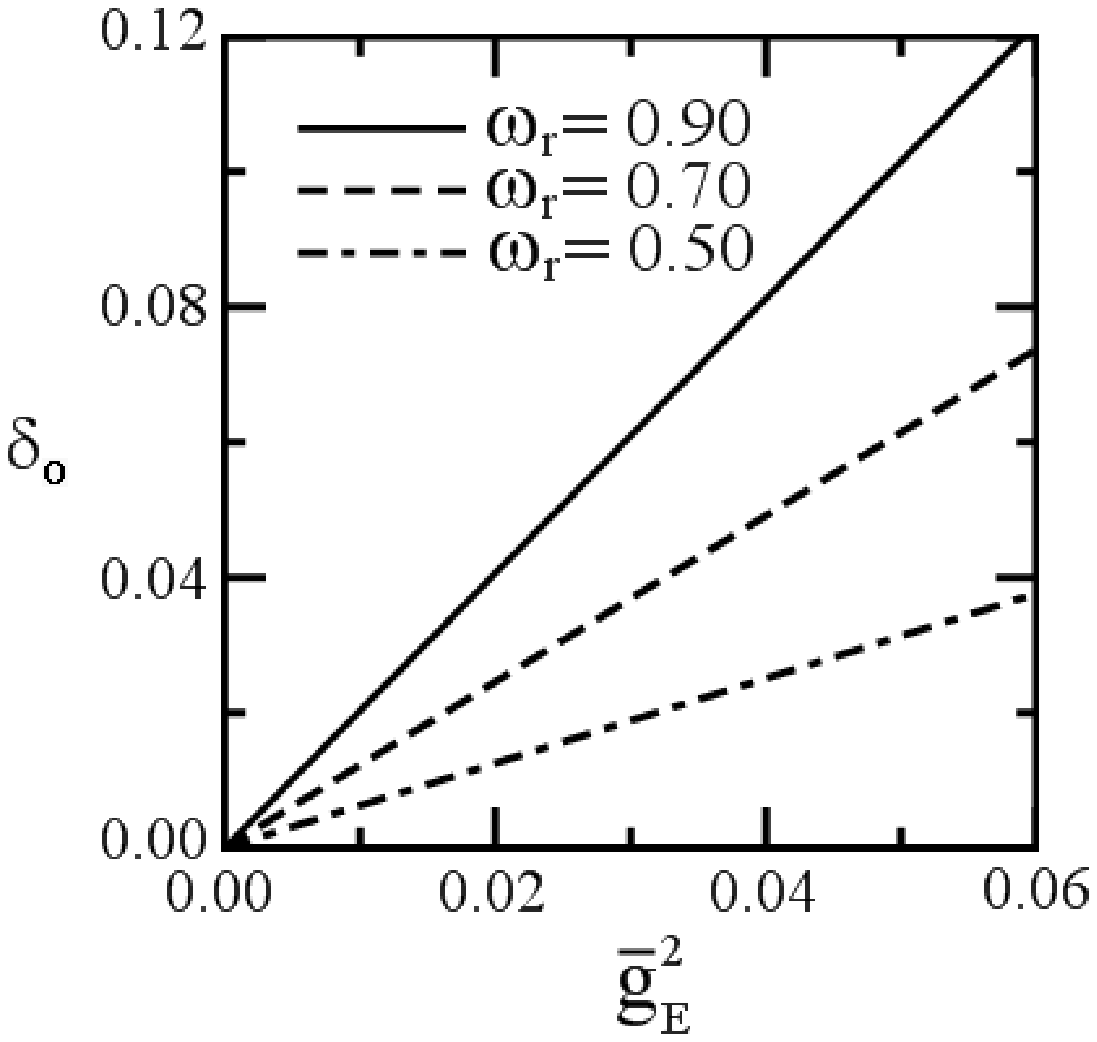}
\caption{The shift $\delta_o$ in the position of the potential minima
is plotted as a function of the junction-cavity coupling strength
${\bar g}^2_E$ for $\omega_r = 0.50$ (dot-dashed line), 0.70 (dashed
line), and 0.90 (solid line).}
\label{deltavsg}
\end{figure}

Another important effect of the resonant cavity is the shift $\delta_o$ in the position of
potential minima.  As the potential barrier height is reduced, the position of the
potential minima are closer together.  The shift $\delta_o$ from the initial position of
$q=\pm q_o/2$ is given by
\begin{equation}
\delta_o = {{\bar g}_E^2 q_o \omega_r^2 \over \epsilon}
{\cosh^2 \ell  \tanh^2 \ell \over {\cosh 2\ell -7}}~.
\end{equation}
Here, we obtained $\delta_o$ by imposing the condition $[dV(q)/dq\vert_{q=(q_o/2)_\pm}=0$,
where $(q_o/2)_\pm = \pm[(q_o/2)-\delta_o]$ denotes the new potential minima.  This shift
$\delta_o$ modifies the constant of motion $E_Q$.  The new value for $E_Q$ may be obtained
from the condition $[dq(\tau)/d\tau]_{(q_o/2)_\pm} = 0$, noting that the fluxon is
initially located at the bottom of either side of the double-well potential so that
$V_Q( (q_o/2)_\pm) = 0$.  We plot the numerically computed shift $\delta_o$ as a function
of ${\bar g}^2_E$ in Fig. \ref{deltavsg} for $\omega_r = 0.50$ (dot-dashed line), 0.70
(dashed line), and 0.90 (solid line) to illustrate the amount of this shift in the
weak-coupling regime.  The curves indicate that $\delta_o$ increases with ${\bar g}_E^2$
and with $\omega_r$, reflecting potential renormalization.

The resonent cavity also modifies the oscillation frequency $\omega_e$ at the potential
minima.  The modified frequency $\omega_e$ is given by
\begin{equation}
\omega_e \approx \omega_o \left\{ 1 + {{\bar g}_E^2 \over M}
\left[ 1 + {\omega_r^2 \over \omega_o^2}(1-\Upsilon) \right] \right\}~,
\end{equation}
where $\omega_o$ is the frequency in the absence of resonant cavity and $\Upsilon = 6q_o
\sinh 2q_o \tanh \ell /\epsilon (\cosh^2\ell -4)\sinh^2 \ell$.

We now combine these effects together and estimate the ground-state energy
splitting\cite{GWH} $\Delta_{cav}$ by using the action $S_{eff}^Q[q]$ of Eq.
(\ref{actionQ}) and by using the standard method of summing over the "instanton"
trajectories.\cite{Cole}   By following Weiss and coworkers,\cite{WGHR} we compute the
one-bounce contribution to the partition function ${\cal Z}_{fluxon}$, assuming that the
fluxon is initially pinned at one of the potential minima.   We write the partition
function as
\begin{equation}
{\cal Z}_{fluxon} = \sum_{i=0}^{\infty} {\cal Z}_i
\end{equation}
where ${\cal Z}_i$ denotes the $i$-bounce contribution.  Here the bounce is an
instanton-anti-instanton pair.  To estimate $\Delta_{cav}$, we compute both the
saddle-point (${\cal Z}_0$) and the one-bounce (${\cal Z}_1$) contribution to
${\cal Z}_{fluxon}$ by noting that ${\cal Z}_1$ may be expressed as
\begin{equation}
{\cal Z}_1 = {{\cal Z}_0 \over 2\pi} \left( {\Delta_{cav}\theta \over 2} \right)^2~,
\label{Z1}
\end{equation}
where $\theta = 1/T$.  For the contribution ${\cal Z}_0$, we assume that the fluxon is
initially confined at $q=(q_o/2)_-$ and obtain
\begin{equation}
{\cal Z}_0= N \left( \prod_{n=0}^\infty \lambda_n^o \right)^{-1/2}
\end{equation}
where the eigenvalues $\lambda_n^o$ are determined from
\begin{eqnarray}
\left[ -M_e \partial_\tau^2 + V_Q^{''}
\left( -{q_o \over 2}+\delta \right) \right] q_n^o(\tau)
~~~~~~~~~~~~~~~~~~~~~~~~~~
\nonumber \\
+ 4\pi {\bar g}_E^2 \int_{-\theta/2}^{\theta/2}
K(\tau-\tau')~q_n^o(\tau') = \lambda_n^o q_n^o(\tau) .~
\end{eqnarray}
Here $\partial_\tau^2 = \partial^2/\partial\tau^2$, $V_Q^{''}(q)=\partial^2 V_Q(q)/
\partial q^2$, and the cavity kernel $K(\tau - \tau') = (\omega_r^3/2)
\exp [-\omega_r \vert \tau - \tau' \vert ]$ accounts for the non-local effect.

For the one-bounce contribution ${\cal Z}_1$ to ${\cal Z}_{fluxon}$, we separate the
center coordinate $q(\tau)$ into two parts as
\begin{equation}
q(\tau) = {\bar q}(\tau) + \sum_{n=0}^\infty c_n q_n(\tau)~,
\label{trasep}
\end{equation}
where ${\bar q}(\tau)$ describes a bounce-like trajectory and the remaining terms describe
the arbitrary paths about this bounce-like trajectory.  This separation of $q(\tau)$ may
be used to write the action $S_{eff}^Q[q]$ as
\begin{equation}
S_{eff}^Q [q(\tau)] = S_{B,1}^{cav}({\bar q}(\tau)) +
\sum_{n=0}^\infty {1 \over 2} \lambda_n c_n^2~.
\end{equation}
Here $S_{B,1}^{cav}$ accounts for the one-bounce-like trajectory in the resonant cavity.
We choose $q_n(\tau)$ of Eq. (\ref{trasep}) so that the eigenfunctions of the second
variational derivative of $S_{eff}^Q [q]$ at ${\bar q}$ and the eigenvalues $\lambda_n$
are determined from
\begin{eqnarray}
\left[ -M_{e} \partial_\tau^2 + V_Q^{''}\left({\bar q}
\right)\right] q_n(\tau)
~~~~~~~~~~~~~~~~~~~~~~~~~~~~~~~~~~~~~
\nonumber \\
~~~~~~
+ 4\pi {\bar g}_E^2 \int_{-\theta/2}^{\theta/2} K(\tau-\tau')~q_n(\tau') =
\lambda_n q_n(\tau).~~
\label{eqigenZ1}
\end{eqnarray}
We note that the first two eigenvalues, $\lambda_0$ and $\lambda_1$, need to be separated
from the rest because $\lambda_0 \leq 0$ and $\lambda_1 = 0$ while the other eigenvalues
are positive.  The one-bounce contribution (${\cal Z}_1$) may be expressed as
\begin{equation}
{\cal Z}_1 = N \int \prod_{n=0}^{\infty} {dc_n \over \sqrt{2\pi}}
e^{-\left ( S_{B,1}^{cav} +
{1 \over 2} \sum_{n=0}^{\infty} \lambda_nc_n^2 \right )}~,
\label{Z1enl}
\end{equation}
where $N$ is a normalization constant.  With the separation of the first two eigenvalues
(i.e., $\lambda_0 \leq 0$ and $\lambda_1=0$) from the others, we write the one-bounce
 contribution to the partition function as
\begin{eqnarray}
{\cal Z}_1 \approx {{\cal Z}_0 \theta \over 2\pi}
\left [ \int_0^{\theta} d\tau_1 e^{-S_{B,1}^{cav}(\tau_1)}
\right] \left[ {\prod_{n=0}^\infty \lambda_n^o \over
\prod_{n=2}^\infty \lambda_n} \right]^{1/2} \times
\nonumber \\
\left [ \int_{-\theta/2}^{\theta/2} d\tau
\left( {d{\bar q} \over d\tau_1} \right)^2 \right]^{1/2}
\left [ \int_{-\theta/2}^{\theta/2} d\tau'
\left( {d{\bar q} \over d\tau'} \right)^2 \right]^{1/2}.~
\label{Z1e}
\end{eqnarray}
We now need to evaluate ${\cal Z}_1$ of Eq. (\ref{Z1e}) to estimate $\Delta_{cav}$.  Using
Eq. (\ref{Z1}), we write the ground-state energy splitting $\Delta_{cav}$ as
\begin{equation}
\Delta_{cav} = {2\omega_e \over \sqrt{\pi}}
\left( R_{cav} L_{cav} e^{-S_{B,1}^{cav}} \right)^{1/2}
\label{split}
\end{equation}
where the dimensionless factors $R_{cav}$ and $L_{cav}$ are
\begin{equation}
R_{cav} = {1 \over M_e\omega_e^2}
\left( {\prod_{n=0}^\infty \lambda_n^o \over
\prod_{n=2}^\infty \lambda_n} \right)^{1/2}
\label{Rcav}
\end{equation}
and
\begin{equation}
L_{cav} ={M_e \over 2} \left [ \int d\tau
\left( {d{\bar q} \over d\tau_1} \right)^2 \right]^{1/2}
\left [ \int d\tau'
\left( {d{\bar q} \over d\tau'} \right)^2 \right]^{1/2} ,
\label{Lcaveq}
\end{equation}
respectively.  The exponent $S_{B,1}^{cav}$ is given by
\begin{equation}
S_{B,1}^{cav} = \int_{-\theta/2}^{\theta/2} d\tau
\left[ {M_e \over 2} \left( {dq(\tau) \over d\tau} \right)^2 +
V_Q(q) \right]~.
\label{actbounce}
\end{equation}
This exponent accounts for the contribution from the two transversal of the potential
barrier.  We note that the exponent $S_{B,1}^{cav}$ of Eq. (\ref{actbounce}) does not
contain the non-local contribution, as in Eq. (\ref{tunrate}), because this contribution
is already included in the calculation of ${\cal Z}_1$ (see Eq. (\ref{Z1enl})).  We now
compute $R_{cav}$, $L_{cav}$ and $S_{B,1}^{cav}$, separately, to determine the
ground-state energy splitting $\Delta_{cav}$.  To focus on the effects due to the
junction-cavity interaction, we present the details of the calculation for $R_{cav}$ and
$L_{cav}$ in Appendix A and B, respectively, and discuss the dependence of these factors
on the junction-cavity coupling strength ${\bar g}_E^2$.

\begin{figure}[t]
\includegraphics[width=6.3cm]{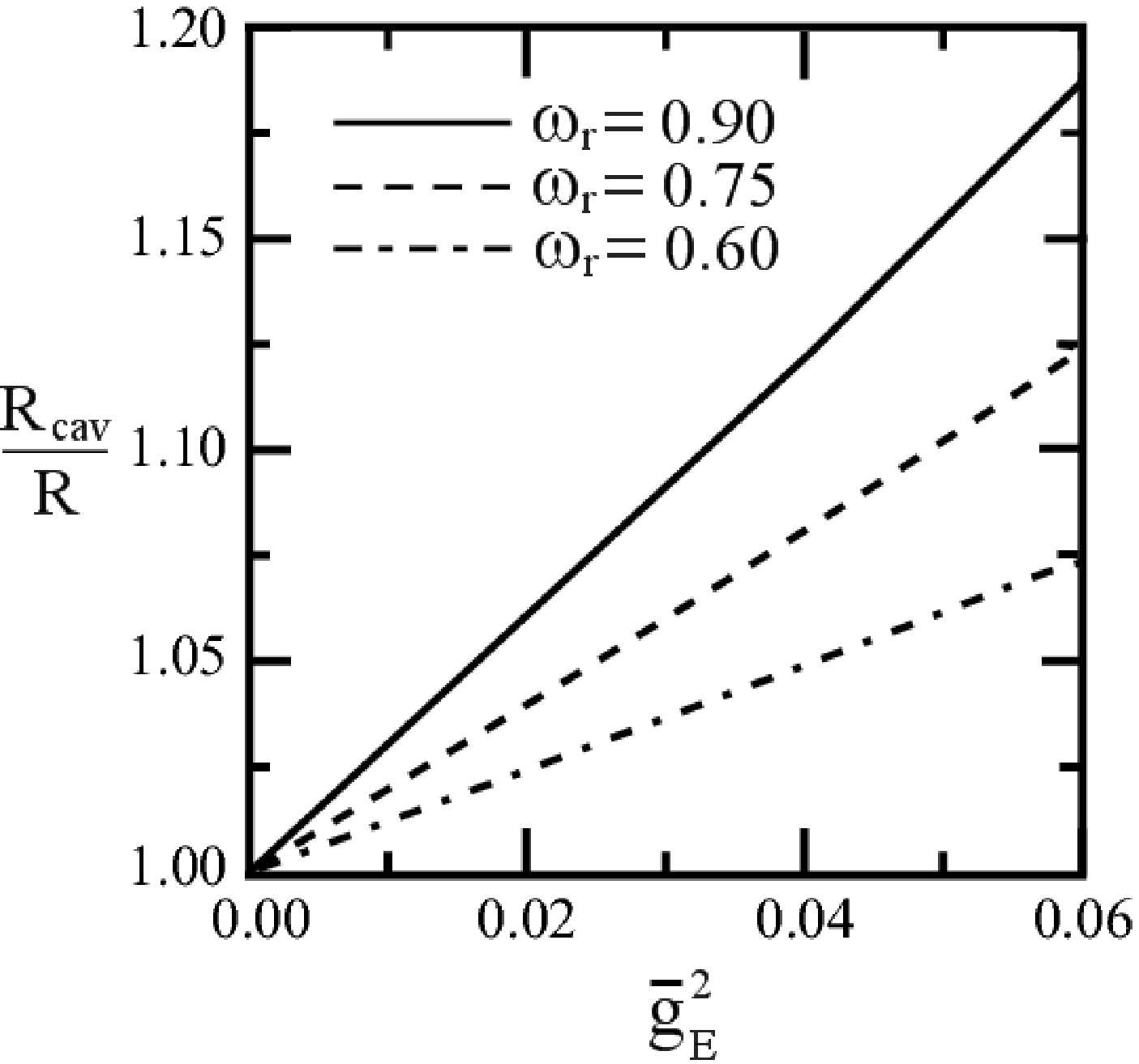}
\caption{The numerically computed ratio of the dimensionless factor
$R_{cav}/R$ is plotted as a function of the junction-cavity coupling
strength ${\bar g}_E^2$ for $\omega_r =$ 0.60 (dot-dashed line), 0.75
(dashed line), and 0.90 (solid line). }
\label{RcavR}
\end{figure}

The dimensionless factor $R_{cav}$ in the weak-coupling regime is given by
\begin{equation}
R_{cav} \cong 2 +  {\pi {\bar g}_E^2 \omega_r^2 \over {2M\omega_o^2}}
{ X_R \over (\omega_r+\omega_o)^3}~,
\label{Rcav1}
\end{equation}
where $X_R =\omega_r^3 + 15\omega_r^2\omega_o + 12\omega_r\omega_o^2 - 2\omega_o^3$.
Equation (\ref{Rcav1}) yields the value $R_{cav} = 2$ in the absence of resonant cavity
(i.e., ${\bar g}_E^2 =0$).\cite{Garg}   In Fig. \ref{RcavR}, we plot the numerically
computed ratio $R_{cav}/R$ as a function of ${\bar g}_E^2$ for $\omega_r =$ 0.60
(dot-dashed line), 0.75 (dashed line), and 0.90 (solid line) to illustrate enhancement of
$R$ due to resonant cavity.  The curves indicate that $R_{cav}/R$ increases from 1 almost
linearly with increasing ${\bar g}_E^2$ and $\omega_r$.

\begin{figure}[t]
\includegraphics[width=6.3cm]{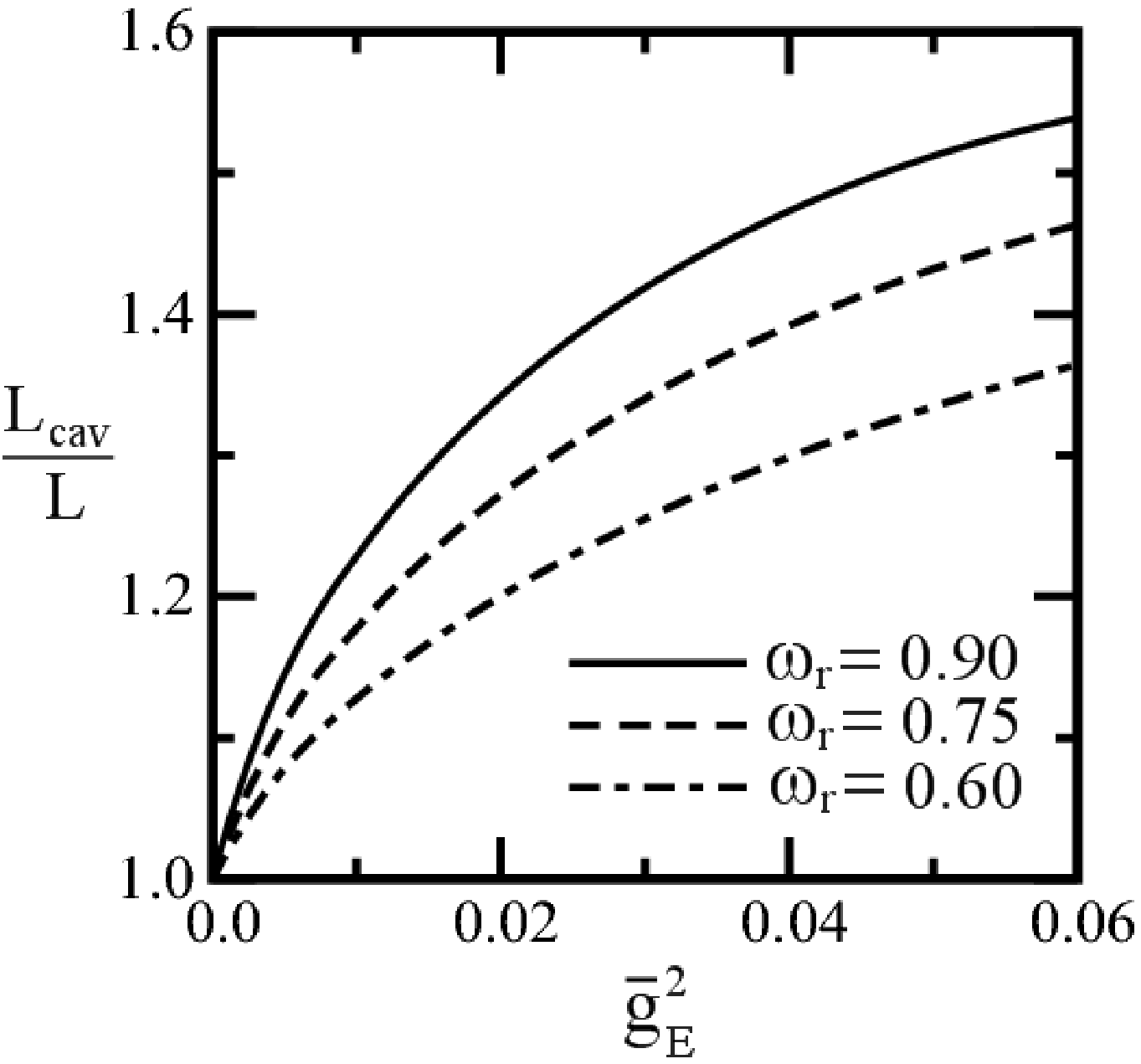}
\caption{The numerically computed ratio of
$L_{cav}/L$ is plotted as a function of the junction-cavity coupling
strength ${\bar g}_E^2$ for $\omega_r = 0.60$ (dot-dashed line), 0.75
(dashed line), and 0.90 (solid line) to illustrate the enhancement. }
\label{LcavLvsg}
\end{figure}

For the dimensionless factor $L_{cav}$, we evaluate the integral of Eq. (\ref{Lcaveq}) by expanding the function $Q(\tau)$ which accounts for the non-local contribution to the
bounce-like trajectory as a power series.  (See Appendix B.)  In the weak-coupling regime
(i.e., ${\bar g}_E^2 \ll 1$), we obtain
\begin{equation}
L_{cav} \approx {\cal V}_M [ A_o
+ {\bar g}_E^2 (B_0 + B_2 q_o^2 + B_4 q_o^4)]~,
\label{Lcavr}
\end{equation}
by retaining the leading order contribution (in ${\bar g}_E^2$).  Here ${\cal V}_M = q_o
\sqrt{2MV_o}$, $A_o= 1 - q_o^2 (2\epsilon b_1/3V_o) - q_o^4 (4\epsilon b_2 / 15 V_o)$,
$B_0 =-(\epsilon + 8b_3\omega_r^2)/8\epsilon$,
$B_2 =[b_1\epsilon + 2(6b_1b_3-1)\omega_r^2 + 2\pi d_1 \omega_r^3]/6V_o$, and
$B_4 = [b_2 \epsilon + 20b_2 b_3\omega_r^2 + \epsilon) + 3\pi d_3\omega_r^3]/15V_o$.  The
frequency independent constants $b_i$ are $b_1 =(\cosh \ell - 2) \rm{sech}^4 (\ell/2)$,
$b_2= (\cosh 2\ell - 26\cosh\ell + 33)/(\cosh\ell + 1)^3$, and
$b_3=(\sinh\ell \tanh \ell)^2/ (\cosh 2\ell - 7)$.  Equation (\ref{Lcavr}) indicates that
$L_{cav}$ in the resonant cavity is larger than $L={\cal V}_M A_o$ in its absence.
However, due to the functional form of $L_{cav}$, the enhancement of $L_{cav}$ from $L$
deviates from the linear dependence on ${\bar g}_E^2$ at a smaller value than that for
$R_{cav}$.  To illustrate this deviation, we numerically compute $L_{cav}$ and plot the
ratio $L_{cav}/L$ in Fig. \ref{LcavLvsg} as a function of ${\bar g}_E^2$ for
$\omega_r = 0.60$ (dot-dashed line), 0.75 (dashed line), and 0.90 (solid line).   The
curves show nonlinear enhancement of the dimensionless factor $L_{cav}$ for much
smaller value of ${\bar g}_E^2$ than that for $R_{cav}$ shown in Fig. \ref{RcavR}.

\begin{figure}[t]
\includegraphics[width=6.3cm]{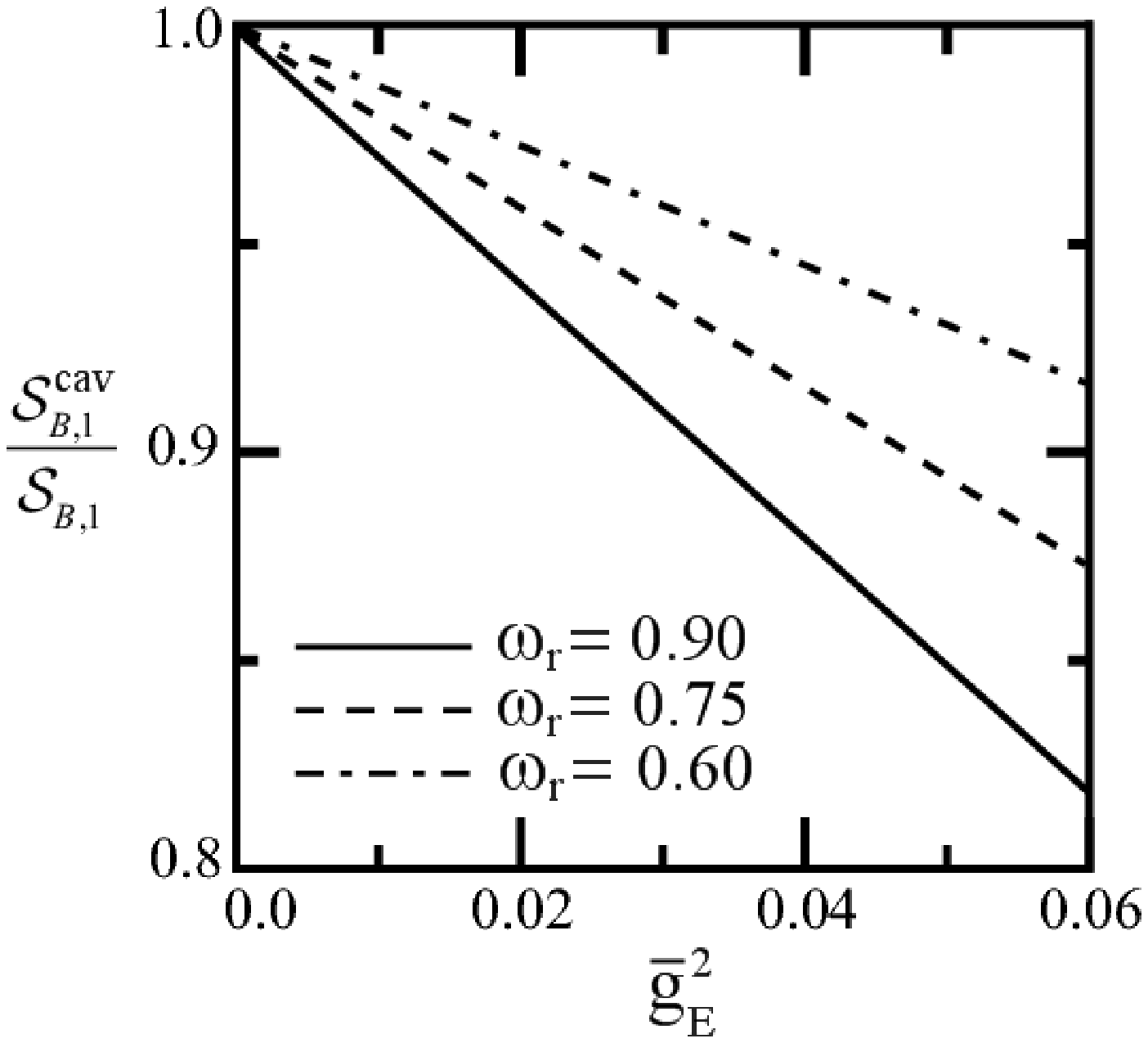}
\caption{The numerically computed ratio of the action $S_{B,1}^{cav}/S_{B,1}$ is plotted
as a function of ${\bar g}_E^2$ for $\omega_r = 0.60$ (dot-dashed line), 0.75 (dashed
line), and 0.90 (solid line) to illustrate that the one-bounce-like action is reduced. }
\label{ScavOverS}
\end{figure}

Finally, we estimate the effects of junction-cavity interaction on $S_{B,1}^{cav}$.  The
action $S_{B,1}^{cav}$ of Eq. (\ref{actbounce}) for the bounce-like trajectory is given by
\begin{equation}
S_{B,1}^{cav} =  2 \int_{(q_o/2)_-}^{(q_o/2)_+}
dq \sqrt{ 2M_e V_Q(q)}~.
\label{onebounce}
\end{equation}
The integral of Eq. (\ref{onebounce}) is evaluated in the same way as that for $L_{cav}$
(see Appendix B).  Again, we simplify the calculation by writing $V_Q$ as a power series
in $q$ and then expand $\sqrt{V_Q(q)}$ in powers of ${\bar g}_E^2$ as
\begin{eqnarray}
\sqrt{V_Q} \approx {{\cal V}_M \over \sqrt{2M_e} q_o} \bigg\{
1 - {2\epsilon q^2 \over V_o}\left(b_1 + {4b_2 \over 3}q^2
+ \cdot\cdot\cdot \right)~~~~~~
\nonumber \\
- {\bar g}_E^2 \left[ {1 \over 8} + {\omega_r^2q_o^2 \over 2V_o} -
{\epsilon q^2 \over 2V_o} \left({\bar b}_1 +
{2  \over 3} b_2q^2 \right) \right] \bigg\},~~~
\end{eqnarray}
where ${\bar b}_1=b_1 + (\omega_r^2/\epsilon)$.  Using this series expansion for
$\sqrt{V_Q}$, we evaluate Eq. (\ref{onebounce}) and obtain $S_{B,1}^{cav}$ to the leading
order in ${\bar g}_E^2$ as
\begin{equation}
S_{B,1}^{cav} \approx 4 {\cal V}_M [A_o
+ {\bar g}_E^2 (B_0 + {\bar B}_2q_o^2 + {\bar B}_4 q_o^4) ]~,
\label{actB1}
\end{equation}
where ${\bar B}_2 = B_2 - (\pi d_1 \omega_r^3/3V_o)$ and ${\bar B}_4 = B_4 -(3\pi d_3
\omega_r^3/15 V_o)$.  The action $S_{B,1}^{cav}$ in the presence of cavity is reduced
from that in its absence (i.e., $S_{B,1}^{cav} < S_{B,1}$).  To illustrate this
suppression of the ratio, we plot the numerically computed ratio $S_{B,1}^{cav}/S_{B,1}$
as a function of ${\bar g}_E^2$ for $\omega_r = 0.60$ (dot-dashed line), 0.75 (dashed
line), and 0.90 (solid line) in Fig. \ref{ScavOverS}.  The curves indicate that in the
one-bounce contribution to the action decreases almost linearly with ${\bar g}_E^2$ in
the weak-coupling region as indicated by Eq. (\ref{actB1}).  This reduction reflects that
the potential barrier height is reduced (see Fig. \ref{VcavvsV}) and the potential minima
become closer together (see Fig. \ref{deltavsg}) with increasing junction-cavity
interaction strength.

\begin{figure}[t]
\includegraphics[width=6.3cm]{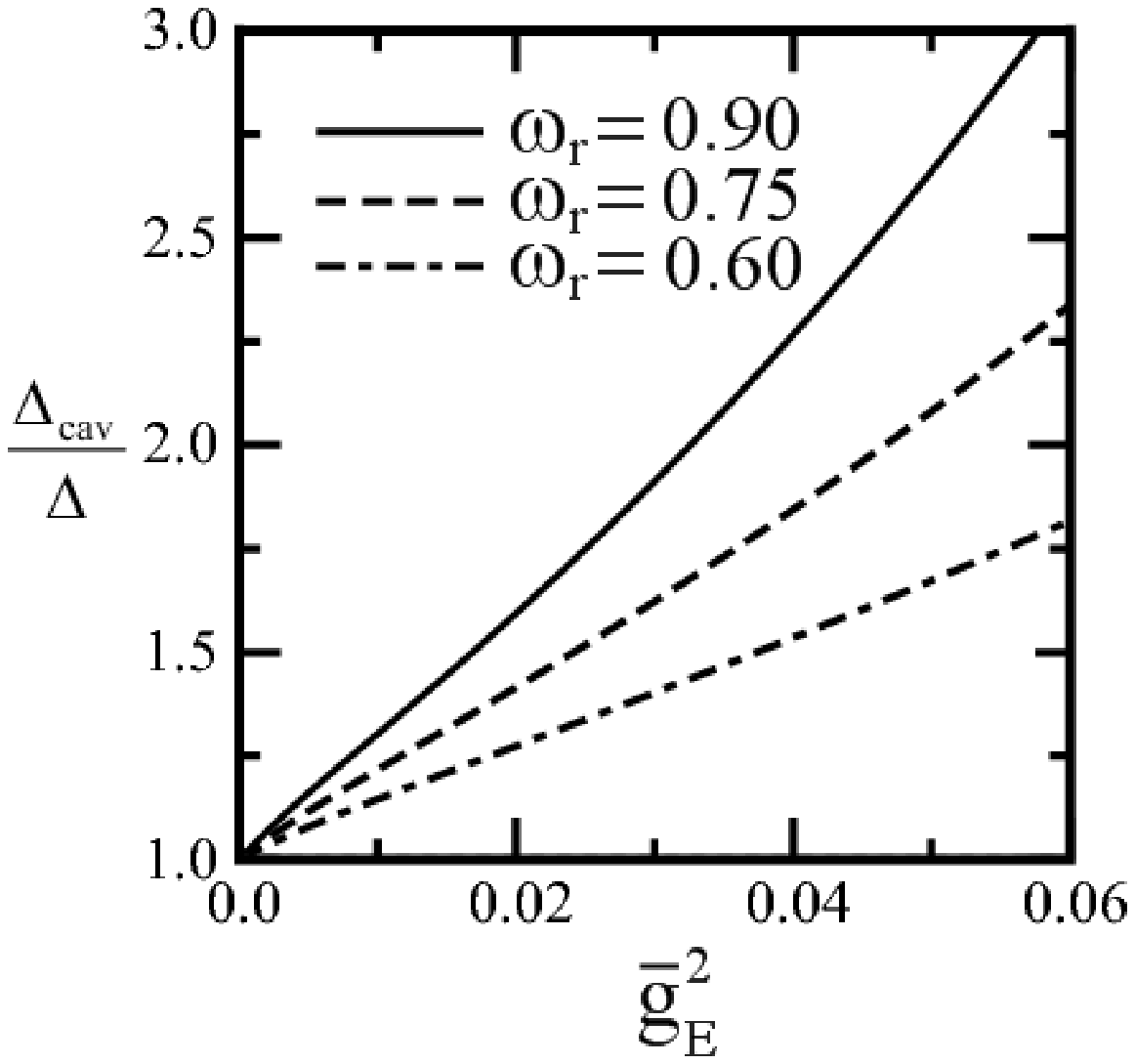}
\caption{The numerically computed ratio of $\Delta_{cav}/\Delta$ is plotted as
a function of the junction-cavity coupling strength ${\bar g}_E^2$ for
$\omega_r = 0.60$ (dot-dashed line), 0.75 (dashed line) and 0.90 (solid line) to
illustrate the enhancement in resonant cavity.}
\label{Delcavdel}
\end{figure}

We now combine the effects of resonant cavity on $R_{cav}$, $L_{cav}$ and $S_{B,1}$
together and estimate the enhancement of the gound-state energy splitting $\Delta_{cav}$
from $\Delta$.  Here, $\Delta$ denotes the energy splitting in the absence of resonant
cavity given by
\begin{equation}
\Delta= 2 {\cal A} \left({S_o \over  2\pi} \right)^{1/2} e^{-S_o} ,
\end{equation}
where ${\cal A}= [{\prod_{n=0}^\infty \lambda_n^o / \prod_{n=1}^\infty \lambda_n}]^{1/2}$
and $S_o$ denotes the action integral.  In the weak-coupling regime, the ratio
$\Delta_{cav}/\Delta$ to the leading order in ${\bar g}_E^2$ is given by
\begin{eqnarray}
{\Delta_{cav} \over \Delta} \approx  1 + {{\bar g}_E^2 \over M}
\bigg\{ 1 &+& {\omega_r^2 \over \omega_o^2} \left[
1-\Upsilon +{\pi X_R \over 8(\omega_r+\omega_o)^3} \right]
\nonumber \\
&+&{M \over 2A_o} (B_0 + B_2 q_o^2 + B_4 q_o^4)
\\
&-& 2M{\cal V}_M (B_0 + {\bar B}_2q_o^2 + {\bar B}_4 q_o^4) \bigg\}.~~~~
\nonumber
\end{eqnarray}
The result indicates that $\Delta_{cav}$ is enhanced with increasing ${\bar g}_E^2$ and
$\omega_r$.  To illustrate this enhancement, we numerically compute and plot
$\Delta_{cav}/\Delta$ as a function of ${\bar g}_E^2$ for $\omega_r = 0.60$ (dot-dashed
line), 0.75 (dashed line) and 0.90 (solid line) in Fig. \ref{Delcavdel}.  The curves show
that $\Delta_{cav}/\Delta$ increases roughly linearly with ${\bar g}_E^2$ from
${\bar g}_E^2=0$ to 0.02.  However, the deviation from this linear behavior becomes
noticeable for ${\bar g}_E^2 \geq 0.02$.  Also, $\Delta_{cav}/\Delta$ increases
significantly from 1 at ${\bar g}_E^2 = 0$ in the weak-coupling regime.  We note that the
corresponding changes in the ratio $R_{cav}/R$, $L_{cav}/L$ and $S_{B,1}^{cav}/S_{B,1}$
over the same range of ${\bar g}_E^2$ are less significant.  For instance,
$\Delta_{cav}/\Delta$ for $\omega_r=0.90$ increases from 1.0 to 1.45 for the increase of
${\bar g}_E^2$ from 0.0 to 0.015.  Over the same range of ${\bar g}_E^2$, $R_{cav}/R$,
$L_{cav}/L$ and $S_{B,1}^{cav}/S_{B,1}$ change from 1.0 to 1.05, from 1.0 to 1.29, and
from 1.0 to 0.96, respectively.  The notable increase in $\Delta_{cav}/\Delta$ compared to
$R_{cav}/R$, $L_{cav}/L$ and $S_{B,1}^{cav}/S_{B,1}$ reflects that $\Delta$ is
small.\cite{KDP}   Hence, $\Delta_{cav}$ depends sensitively on the variation of the
exponent $S_{B,1}^{cav}$.

\section{summary and conclusion}

In summary, we investigated the effects of high-$Q_c$ resonant cavity on MQT of fluxon
from metastable state in a single LJJ and in a stack of two coupled LJJs.  Also, we
estimated the ground-state energy splitting for fluxon in a double-well potential.  We
find that both the tunneling rate and the ground-state energy splitting are increased in
the resonant cavity.  However, the amount of these increases is significantly different.
For MQT of the fluxon, the tunneling rate increases due to the renormalization of fluxon
mass, but negligible in the weak-coupling regime.  On the other hand, the increase in the
ground-state energy splitting is due to potential renormalization, but this increase can
become significant with increasing ${\bar g}_E^2$ as shown in Fig. \ref{Delcavdel}.  This
energy splitting enhancement is consistent with the result of increase in the energy
separation due to the interaction between a two-level system and a quantized radiation
field, described by the Jaynes-Cummings (JC) model.\cite{JC}  Moreover, the consistency\cite{QuanOpt} between the result of the present work and that of the JC model
indicates that the effective Hamiltonian for the JVQ-cavity system may be similar to the
JC model.

The effects due to i) interaction between the JVQ and a dissipative environment and ii)
the losses resulting from a low-Q cavity are neglected in the present work.
These dissipative effects are expected to be present in real systems and may be
accounted by using an effective spectral density which characterizes the form of
dissipation.\cite{GOA}  Inclusion of both the dissipative environment and cavity losses
may reduce the size of increase in the ground-state energy splitting and may lead to
decrease in the energy spliting when the dissipative effects become strong, as indicated
by the analysis of dissipative two-state systems.\cite{LCDFGZ}  However, these dissipation
contributions do not reverse the effects due to the potential renormalization completely
in weakly dissipative systems.

Enhancement of ground-state energy splitting due to the junction-cavity interaction
may have an important consequence for the decoherence time of JVQ in the resonant cavity.
Earlier study\cite{KDP} of the JVQ decoherence time by Kim, Dhungana and Park indicates
that the increase in the decoherence time in noisy environment (i.e., $T_{\phi}^{noise}$)
is correlated with the increasing ground-state energy splitting $\Delta$.  This suggests
that, as $\Delta$ may be tuned by adjusting the strength of junction-cavity interaction,
the resonant cavity may be used to control the property of JVQ.  For instance, the
decoherence time $T_{\phi}^{noise}$ may be increased by increasing the strength of
interaction between fluxon and cavity EM mode.   Also, due to the similarities between a
cavity EM mode and an optical phonon mode, the interaction between fluxon and optical
phonons in the LJJ may affect the decoherence time.

Another important property of JVQs is entanglement between the qubits.  As our result
suggests that the decoherence time for JVQ can be increased by increasing the strength of
junction-cavity interaction, the resonant cavity may also be useful for tuning the level
of entanglement between the JVQs.  Our study suggests that the present approach for JVQs
is similar to the microwave cavity approach used for the other superconductor
qubits.\cite{cQED}  The effective Hamiltonian for the multiple JVQs in a resonant cavity
may resemble the Tavis-Cummings model\cite{TC} which is the extension of the JC model to the case of multiple qubits.  This similarity may be exploited by using the resonant
cavity to control the level of concurrence\cite{conc} for JVQs since the junction-cavity
interaction may also promote entanglement.  Hence, the effects of resonant cavity on
entanglement between the interacting JVQs would be an interesting area for further study.

\vskip 0.3in

The authors would like to thank W. Schwalm and K.-S. Park for helpful discussions and
I. D. O'Bryant for assisting with part of the numerical calculation.

\vskip 0.3in

\centerline{ {\bf APPENDIX A:}}

\centerline{ CALCULATION OF $R_{cav}$}
$\\$
For convenience, the dimensionless factor $R_{cav}$ of Eq. (\ref{Rcav}) is estimated in
the continuum limit.  In this limit, we may write $R_{cav}$ as
\begin{equation}
R_{cav} = \exp\bigg\{ {1\over \pi} \int_{M_e\omega_e^2}^\infty
{d\lambda \over \lambda} [\delta_+(\lambda)+ \delta_-(\lambda)]
\bigg\}~,
\label{AppRcav}
\end{equation}
where $\delta_\pm(\lambda)$ denotes the phase shift due to the scattering potential $U$.
This phase shift may be expressed as
\begin{equation}
\delta_\pm(\lambda )=\cot^{-1} \left [
{U^{-1} -g_\lambda'(0)\mp g_\lambda'(\tau_s) \over
g_\lambda'' (0) \pm g_\lambda'' (\tau_s)} \right]~,
\label{phases}
\end{equation}
where $\tau_s=-\theta/2$, and $g_\lambda' (\tau)$ and $g_\lambda'' (\tau)$ denote the real
and imaginary part of the Green's function (i.e., $g_\lambda(\tau) = g_\lambda' (\tau )
+ i g_\lambda'' (\tau )$).  The phase shift $\delta_\pm (\lambda)$ due to the scattering
from the net potential difference of
\begin{eqnarray}
V_Q''({\bar q}) - V_Q''((q_o/2)_-) =
~~~~~~~~~~~~~~~~~~~~~~~~~
\nonumber \\
~~~~~~~~~~~~~~~~~~
-U \left[ \delta \left(\tau + {\tau_s \over 2}\right)
+ \delta \left(\tau - {\tau_s \over 2}\right) \right]
\end{eqnarray}
consists of two Dirac $\delta$-functions at $\tau = \pm \tau_s/2$. The strength of the
scattering potential $U$ is given by
\begin{equation}
U^{-1} = g_0(0) - g_0(\tau_s)~,
\end{equation}
where $g_0(\tau )$ is the Green's function for the eigenvalue $\lambda=0$.  The Green's
function $g_\lambda (\tau)$ is written as
\begin{equation}
g_\lambda(\tau) =\int_{-\infty}^\infty {d\omega \over 2\pi}
{e^{i\omega\tau} \over  M_e[\omega^2 + \zeta(\omega ) +
\omega_e^2] - \lambda -i\delta}~.
\end{equation}
Here the effects of the resonant cavity are accounted for via $M_e$, $\omega_e$ and
$\zeta(\omega)$.  The function $\zeta(\omega)$, obtained from the cavity kernel $K(\tau)$
of Eq. (\ref{kerqubit}),
\begin{equation}
\zeta (\omega) ={4\pi {\bar g}_E^2 \over M_e} {\omega_r^4 \over
\omega^2 + \omega_r^2} ,
\end{equation}
reflects that the resonant cavity supports a single-mode with frequency $\omega_r$.  Using
the function $\zeta(\omega)$, we write the real part of the Green's function as
$g_\lambda'(\tau)=g_{\lambda,+}'(\tau) + g_{\lambda,-}'(\tau)$, where
\begin{equation}
g_{\lambda,\pm}'(\tau)={-1 \over 4 M_e \omega_{\lambda,\pm}}
\left( 1 \pm {{\omega_r^2 + \omega_{1,\lambda}^2} \over
2\omega_{2,\lambda}^2} \right)
\sin \omega_{\lambda,\pm} \tau~,
\label{greal}
\end{equation}
$\omega_{\lambda,\pm} = (\omega_{1,\lambda}^2 \pm \omega_{2,\lambda}^2)^{1/2}$,
$\omega_{1,\lambda}^2 = [(\lambda/M_e) - \omega_e^2 - \omega_r^2 ]/2$, and
$\omega_{2,\lambda}^2 = \{[(\lambda/M_e) - \omega_e^2 + \omega_r^2]^2 -
(16\pi g_E^2/ M_e) \omega_r^4 \}^{1/2}$.  On the other hand, we write the imaginary part
of the Green's function as $g_\lambda''(\tau) = g_{\lambda,+}''(\tau) +
g_{\lambda,-}''(\tau)$, where
\begin{equation}
g_{\lambda,\pm}''(\tau)={1 \over 4 M_e \omega_{\lambda,\pm}}
\left( 1 \pm {\Omega^2 + \omega_{1,\lambda}^2 \over
2\omega_{2,\lambda}^2} \right) \cos \omega_{\lambda,\pm} \tau~.
\end{equation}
We note that the phase shift $\delta_\pm(\lambda)$ has both slowly varying and rapidly
oscillating contributions.  For an extended bounce (i.e., $\omega_e \tau_s \gg 1$), the
rapidly oscillating terms become negligible compared to the non-oscillating terms.

The factor $R_{cav}$ of Eq. (\ref{AppRcav}) may be simplified by using the substitution
$\lambda = M_e \omega_e^2 (1+p^2)$, where $p$ is a dimensionless momentum variable.  With
this change of variable, we write $R_{cav}$ as
\begin{equation}
R_{cav} = \exp\bigg\{{1\over \pi} \int_0^\infty
{p~dp \over {1+p^2}} [\delta_+(p)+ \delta_-(p)]\bigg\}~.
\label{rcav}
\end{equation}
The factor $R_{cav}$ of Eq. (\ref{rcav}) may be further simplified by neglecting the
rapidly oscillating contributions in the phase shift $\delta_\pm (\lambda)$ of Eq.
(\ref{phases}).  Neglecting these oscillatory contributions, we approximate
$\delta_\pm (p)$ to a simpler form $\delta (p)$ and write the factor $R_{cav}$ as
\begin{equation}
R_{cav} = \exp\bigg\{{2\over \pi} \int_0^\infty
{p~dp \over {1+p^2}} \delta(p)\bigg\}~.
\label{rcava}
\end{equation}
The simplified phase shift $\delta (p)$ is given by
\begin{equation}
\delta(p) = \cot^{-1} \left[ {{U^{-1} - g_p'(0)} \over
{g_p''(0)}} \right]~,
\label{sphases}
\end{equation}
where the scattering potential strength $U$ is given by
\begin{equation}
U^{-1} = {1 \over 4M_e} \left({1-W_0 \over
\sqrt{\omega_{2,0}^2-\omega_{1,0}^2}} +
{1+W_0 \over \sqrt{\vert \omega_{2,0}^2 +
\omega_{1,0}^2 \vert }} \right)~,
\label{uu}
\end{equation}
and $W_0=(\omega_r^2+\omega_{1,0}^2)/\omega_{2,0}^2$.  We note that $\omega_{1,0}$ and
$\omega_{2,0}$ are obtained from $\omega_{1, \lambda}$ and $\omega_{2,\lambda}$ of Eq.
(\ref{greal}) for the eigenvalue  $\lambda=0$, respectively.  The real and imaginary part
of the Green's function are given, respectively, by
\begin{equation}
g_p'(0)= {1 \over 4M_e}{1 - W_p \over
\sqrt{\omega_{2,p}^2 - \omega_{1,p}^2}}
\label{realgp}
\end{equation}
and
\begin{equation}
g_p''(0)= {1 \over 4M_e}{1 + W_p \over
\sqrt{\omega_{2,p}^2 + \omega_{1,p}^2}}~,
\label{imaggp}
\end{equation}
where $W_p=(\omega_r^2+\omega_{1,p}^2)/\omega_{2,p}^2$.  We note that $\omega_{1,p}$ and
$\omega_{2,p}$ are obtained from $\omega_{1, \lambda}$ and $\omega_{2,\lambda}$ of Eq.
(\ref{greal}), respectively, by setting $\lambda=M_e\omega_e^2(1+p^2)$.

We now compute $R_{cav}$ to the leading order in ${\bar g}_E^2$ to account for the effects
of resonant cavity in the weak coupling regime (i.e., ${\bar g}_E^2 \ll 1$).  For this
calculation, we write the renormalized mass of the fluxon as $M_e = M -2{\bar g}_E^2$ and
express the oscillation frequency $\omega_e$ as
\begin{equation}
\omega_e^2 \cong \omega_o^2 \left\{ 1+ {2{\bar g}_E^2 \over M}
\left[ 1 + {\omega_r^2 \over \omega_o^2} (1-\Upsilon) \right] \right\}~.
\end{equation}
Also we rewrite the strength of the potential $U$ as
\begin{equation}
U^{-1} \cong {1 \over 2M\omega_o} - {2\pi {\bar g}_E^2 \omega_r^3
\over M^2(\omega_r^2 -\omega_o^2)^2} \left( 1 -
{X_u \over 32\pi\omega_r^3 \omega_o^3} \right)
\label{uug}
\end{equation}
where $X_u= M\omega_o^2(\omega_o^2 -\omega_r^2)^2
-8[\omega_r^2\Upsilon(\omega_o^2 -\omega_r^2)^2 + 2\pi \omega_r^4 (\omega_r^2 -
3\omega_o^2)]$.  By combining these expressions together, we rewrite the real and imaginary
part of the Green's function of Eqs. (\ref{realgp}) and (\ref{imaggp}), respectively, as
\begin{equation}
g_p'(0) \cong -{{\bar g}_E^2 \pi \omega_r^3 \over
2M^2 (p^2\omega_o^2 + \omega_r^2)^2}
\end{equation}
and
\begin{equation}
g_p''(0)= {1 \over 2Mp\omega_o} \left[ 1 +
{{\bar g}_E^2 X_g \over
8Mp^2\omega_o^2(p^2\omega_o^2 + \omega_r^2)^2} \right]~,
\label{imaggpg}
\end{equation}
where $X_g = 4\omega_r^2[\pi\omega_r^2(3p^2\omega_o^2 + \omega_r^2) - 2 \Upsilon p^2
(p^2\omega_o^2 + \omega_r^2)^2] + Mp^2\omega_o^2(p^2\omega_o^2+\omega_r^2)^2$.  Now, we use
Eqs. (\ref{uug}) - (\ref{imaggpg}) and rewrite the simplified phase shift $\delta (p)$ of
Eq. (\ref{sphases}) as
\begin{equation}
\delta(p) \cong \cot^{-1}p - {{\bar g}_E^2 \pi\omega_r^3 X_p
\over 2Mp(1+p^2)X_\omega}
\label{sphasess}
\end{equation}
where $X_p = - (\omega_o^2\omega_r^3 + 2\omega_o\omega_r^4 + \omega_r^5)
+ p^2(2\omega_o^5 + \omega_o^4\omega_r -4\omega_o^3\omega_r^2
-3\omega_o^2\omega_r^3 - 8 \omega_o\omega_r^4 -4\omega_r^5) - p^4
(16\omega_o^3\omega_r^2 + 8\omega_o^2\omega_r^3) - p^6 (8\omega_o^5 +
4 \omega_o^4\omega_r)$ and $X_\omega =\omega_o^2(\omega_o+\omega_r)^2
(p^2\omega_o^2+\omega_r^2)^2$.  Finally, we substitute $\delta(p)$ of Eq. (\ref{sphasess})
into $R_{cav}$ of Eq. (\ref{rcava}) and evaluate the integral to obtain
\begin{equation}
R_{cav} \cong 2 +  {\pi {\bar g}_E^2 \omega_r^2 \over {2M\omega_o^2}}
{X_R \over (\omega_r+\omega_o)^3}~,
\label{rrcav}
\end{equation}
where $X_R =5\omega_r^3 + 15\omega_r^2\omega_o + 12\omega_r\omega_o^2 - 2\omega_o^3$.
Equation (\ref{rrcav}) yields $R_{cav}=2$ in the absence of the resonant cavity (i.e.,
${\bar g}_E^2=0$) as expected.\cite{Garg}

$\\$

\centerline{ {\bf APPENDIX B:}}
\centerline{ CALCULATION OF $L_{cav}$}
$\\$
The factor $L_{cav}$ of Eq. (\ref{Lcaveq}) may be estimated by determining the bounce-like
trajectories $q(\tau)$.  The trajectories obey the equation of motion given by
\begin{eqnarray}
-M_e {d^2q(\tau) \over d\tau^2} &+& {dV_Q(q) \over dq}
\nonumber \\
&+& 4\pi {\bar g}_E^2
\int_{-\infty}^{\infty} d\tau' K(\tau - \tau') q(\tau) = 0 ~.~~~
\label{inteqL}
\end{eqnarray}
We rewrite the equation of motion in a convenient form by integrating Eq. (\ref{inteqL})
by parts and obtain
\begin{eqnarray}
-{M_e \over 2} \left( {dq \over d\tau} \right)^2 + V_Q(q)
~~~~~~~~~~~~~~~~~~~~~~~~~~~~~~~~~~~~~~~~~
\nonumber \\
+ 4\pi {\bar g}_E^2 \int_{-\infty}^{\infty} d\tau' K(\tau-\tau')
q(\tau) q(\tau') = 0 .~~~
\end{eqnarray}
Using this result, we write the factor $L_{cav}$ as
\begin{eqnarray}
L_{cav} &\approx& {M_e \over 2} \int d\tau
\left ( {dq \over d\tau} \right)^2
\nonumber \\
&=& \int_{(q_o/2)_-}^{(q_o/2)_+} dq
\sqrt{ V_Q(q) + 2\pi {\bar g}_E^2 \omega_r^2 q Q(\tau)}
\label{Lcavsq}
\end{eqnarray}
where $q=q(\tau)$ and
\begin{equation}
Q(\tau) = \int_{-\infty}^{\infty} d\tau'
e^{-\omega_r \vert \tau-\tau'\vert} q(\tau')~.
\label{kerF}
\end{equation}
Here, the non-local contribution due to resonant cavity is accounted for by $Q(\tau)$.  As
discussed in Appendix C, the function $Q(\tau)$ is similar to $q(\tau)$.  By exploiting
this similarity, we expand $Q(\tau)$ in a power series as
\begin{equation}
Q(\tau) = \sum_{n=0}^\infty d_{2n+1} q^{2n+1}(\tau)~,
\label{kernel}
\end{equation}
where $d_{2n+1}$ is the expansion coefficients (see Appendix C).  The power series
expansion for $Q(\tau)$ allows us to evaluate the factor $L_{cav}$ straightforwardly.  By
using this power series expansion, we evaluate the integral of Eq. (\ref{Lcavsq}) in the
weak-coupling regime (i.e., ${\bar g}_E^2 \ll 1$) and obtain the factor $L_{cav}$ to the
leading order in ${\bar g}_E^2$ as
\begin{equation}
L_{cav} \approx {\cal V}_M [ A_o
+ {\bar g}_E^2 (B_0 + B_2 q_o^2 + B_4 q_o^4)]~,
\label{Lcav}
\end{equation}
where ${\cal V}_M = q_o \sqrt{2MV_o}$, $A_o= 1 - q_o^2 (2\epsilon b_1/3V_o) -
q_o^4 (4\epsilon b_2 / 15 V_o)$, $B_0 =-(\epsilon + 8b_3\omega_r^2)/8\epsilon$,
$B_2 =[b_1\epsilon + 2(6b_1b_3-1)\omega_r^2 + 2\pi d_1 \omega_r^3]/6V_o$, and
$B_4 = [b_2 \epsilon + 20b_2 b_3\omega_r^2 + \epsilon) + 3\pi d_3\omega_r^3]/15V_o$.  The
frequency independent constants $b_i$ are given by $b_1 =(\cosh \ell - 2) \rm{sech}^4
(\ell/2)$, $b_2= (\cosh 2\ell - 26\cosh\ell + 33)/(\cosh\ell + 1)^3$, and
$b_3=(\sinh\ell \tanh \ell)^2/ (\cosh 2\ell - 7)$.

$\\$

\centerline{{\bf APPENDIX C:}}

\centerline{POWER SERIES EXPANSION OF $Q(\tau)$}
$\\$
\begin{figure}[t]
\includegraphics[width=6.3cm]{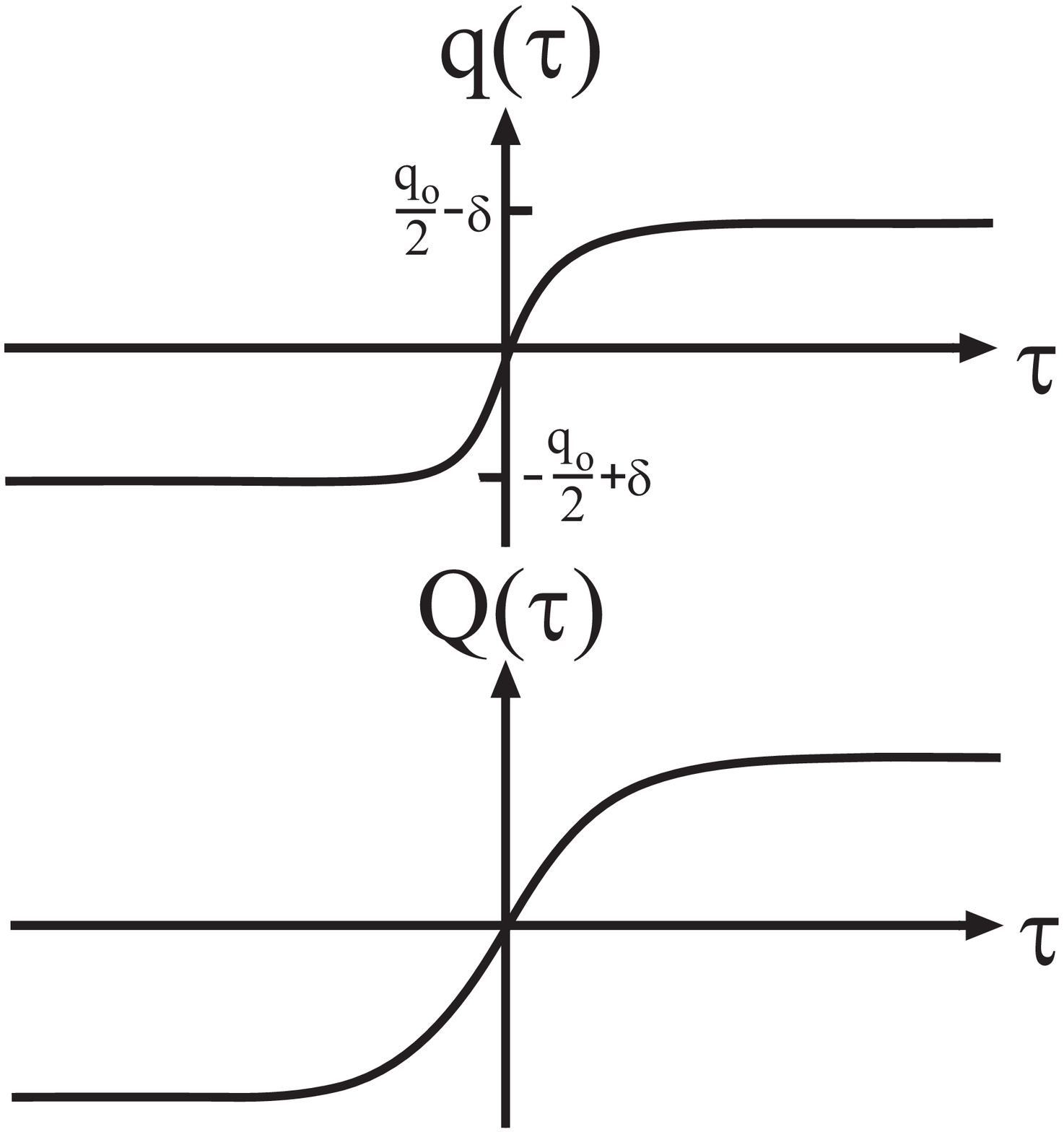}
\caption{Similarity between the function $Q(\tau)$ of Eq. (\ref{kerF}) and the instanton
solution $q(\tau)$ representing the trajectory of the fluxon from one potential minimum to
the other via tunneling is illustrated schematically.  }
\label{instanton}
\end{figure}
The numerically computed function $Q(\tau)$ of Eq. (\ref{kerF}) indicates that $Q(\tau )$
is similar to the functional form of the bounce-like trajectory $q(\tau)$. This similarity
suggests that $Q(\tau )$ is a scaled function of $q(\tau )$ as shown schematically in Fig.
\ref{instanton}.  In this case, we may express the function $Q(\tau)$ as a power series in
$q(\tau)$ as
\begin{equation}
Q(\tau) = \sum_{n=0}^\infty d_{2n+1} q^{2n+1}(\tau)~,
\label{kerd}
\end{equation}
where $d_{2n+1}$ denotes the coefficient for this power series expansion.  We compute the
coefficients $d_{2n+1}$ by starting with a series expansion of $q(\tau)$ in $\tau$ as
\begin{equation}
q(\tau)=\sum_{n=0}^\infty a_{2n+1} \tau^{2n+1}~,
\label{qexpand}
\end{equation}
noting that the instanton solution $q(\tau)$ is an odd function of $\tau$.  Here, the
coefficient $d_{2n+1}$ is obtained by following the five steps as discussed below.
First, we write the bounce-like trajectory $q$ in the absence of resonant cavity.  This
trajectory $q$ may be expressed as
\begin{equation}
q = -b_1 \tau + b_2 \tanh^{-1}(b_3 \tanh q) ,
\label{bounce}
\end{equation}
where the constants $b_1=2\sqrt{\epsilon/M}\coth\ell$, $b_2=(\cosh 2q_o + \cosh \ell)/
\sinh 2q_o$, and $b_3 = \coth q_o$ depend on the parameters $\ell$ and $\epsilon$.  Second,
we expand the right hand side of Eq. (\ref{bounce}) as a power series in $q$ as
\begin{equation}
q = -b_1 \tau + b_2 b_3 q\left( 1 - {{1-b_3^2} \over 3}q^2 +
{{2-5b_3^2+3b_3^4} \over 15}q^4 + \cdot\cdot\cdot \right)~.
\label{bexpand}
\end{equation}
Here, we find the coefficients $a_{2n+1}$ by substituting the series expansion for $q(\tau)$
of Eq. (\ref{qexpand}) into Eq. (\ref{bexpand}).  The first three coefficients
are given by
\begin{eqnarray}
a_1 &=& {b_1 \over b_2 b_3 -1}~,
\nonumber \\
a_3 &=& {b_1^3b_2 b_3(1-b_3^2) \over 3(b_2 b_3 -1)^4}~,
\nonumber \\
a_5 &=& {b_1^5b_2b_3(1-b_3^2)[b_2b_3(3-2b_3^2)+(2-3b_3^2)] \over
15(b_2b_3-1)^7}~.
\nonumber
\end{eqnarray}
Third, we use Eqs. (\ref{kerqubit}) and (\ref{qexpand}) to evaluate $Q(\tau)$ of Eq.
(\ref{kerF}) explicitly as
\begin{equation}
Q(\tau) = \sum_{n=0}^\infty a_{2n+1}
\int_0^\infty d\tau'~e^{-\omega_r \vert \tau-\tau' \vert} \tau'^{2n+1}~.
\label{kerFint}
\end{equation}
Fourth, we evaluate the integrals of Eq. (\ref{kerFint}) and write $Q(\tau)$ in a power
series in $\tau$ as
\begin{eqnarray}
Q(\tau) \approx {2 \over \omega_r} \bigg [
 \tau \left( a_1 + {6a_3 \over \omega_r^2} + {120a_5 \over \omega_r^4}
 \right) ~~~~~~~~~~~~~~
\nonumber \\
~~~~~~~~~~~~
+ \tau^3 \left( a_3 + {20a_5 \over \omega_r^2} \right)
+ \tau^5 a_5  + \cdot\cdot\cdot \bigg]~.
\label{ft1}
\end{eqnarray}
Finally, we use the power series expansion for $q(\tau)$ of Eq. (\ref{qexpand}) and rewrite
$Q(\tau)$ of Eq. (\ref{kerd}) as
\begin{eqnarray}
Q(\tau) = \tau(d_1a_1) + \tau^3(d_1a_3+d_3a_1^3)~~~~~~~~~~~~~~~
\nonumber \\
~~~~~~~
+ \tau^5 (d_1a_5 + 3d_3a_1^2a_3 + d_5a_1^5) + \cdot\cdot\cdot
\label{ft2}
\end{eqnarray}
This series expansion allows us to obtain the expansion coefficients $d_{2n+1}$ by
comparing the power series $Q(\tau)$ of Eqs. (\ref{ft1}) and (\ref{ft2}).  The first three
expansion coefficients, $d_{2n+1}$, are the following:
\begin{eqnarray}
d_1 &=& {2 \over \omega_r}\left(
1+ {6 \over \omega_r^2} {a_3 \over a_1} + \cdot\cdot\cdot \right)~,
\nonumber \\
d_3 &=& {4 \over \omega_r^3}\left[
\left( {10a_5 \over a_1^3} -
{3a_3^2 \over a_1^4} \right) + {60 \over \omega_r^4}
\left( {7a_7 \over a_1^3} - {a_3a_5 \over a_1^4} \right)
+ \cdot\cdot\cdot \right]~,
\nonumber\\
d_5 &=& {12 \over \omega_r^3}\left[
\left( {7a_7 \over a_1^5} -
{11a_3a_5 \over a_1^6} + {3a_3^3 \over a_1^7}\right)
+ \cdot\cdot\cdot \right]~.
\nonumber
\end{eqnarray}
In Sec. V, we use these expansion coefficients to estimate the dimensionless factor
$L_{cav}$ and the one-bounce contribution to the action (i.e., $S_{B,1}^{cav}$).

\end{document}